\documentclass[fleqn,usenatbib]{mnras}

\usepackage{newtxtext,newtxmath}

\usepackage[T1]{fontenc}

\DeclareRobustCommand{\VAN}[3]{#2}
\let\VANthebibliography\thebibliography
\def\thebibliography{\DeclareRobustCommand{\VAN}[3]{##3}\VANthebibliography}

\usepackage{graphicx}
\usepackage{amsmath}	
\usepackage{hyperref}
\usepackage{caption}
\usepackage{bm}
\usepackage{comment}
\usepackage{here}
\usepackage[dvipsnames]{xcolor}
\usepackage{multicol}
\usepackage{lscape}

\title[beaming fraction]{Constraining the Pulsar Beaming Fraction with TeV-Selected Galactic Pulsar Wind Nebulae and unidentified TeV Sources}

\author[T. Shimasue et al.]{
Takumi Shimasue$^{1,2}$\thanks{E-mail: shimasue-takumi@resceu.s.u-tokyo.ac.jp}, Shota Kisaka$^{3,4}$, Aya Bamba$^{1,2,5}$
, Shinpei Shibata$^{6,7}$
\\
$^{1}$Department of Physics, Graduate School of Science, The University of Tokyo,
7-3-1 Hongo, Bunkyo, Tokyo 113-0033, Japan\\
$^{2}$Research Center for the Early Universe, School of Science, The University of Tokyo,
7-3-1 Hongo, Bunkyo, Tokyo 113-0033, Japan\\
$^{3}$Physics Program, Graduate School of Advanced Science and Engineering, Hiroshima University,
Higashi-Hiroshima, Hiroshima, 739-8526, Japan\\
$^4$ Hiroshima Astrophysical Science Center, Hiroshima University, Higashi-Hiroshima, Hiroshima 739-8526, Japan\\
$^{5}$Trans-Scale Quantum Science Institute, The University of Tokyo, Tokyo 113-0033, Japan\\
$^{6}$Faculty of Science, Yamagata University, 
1-4-12 Kojirakawa, Yamagata 990-8560, Japan\\
$^{7}$Institute for Cosmic Ray Research, The University of Tokyo, 
5-1-5 Kashiwanoha, Kashiwa, Chiba, 277-8582, Japan
}
\date{Accepted XXX. Received YYY; in original form ZZZ}

\pubyear{2025}

\begin{document}
\label{firstpage}
\pagerange{\pageref{firstpage}--\pageref{lastpage}}
\maketitle

\begin{abstract}
The pulsar beaming fraction is a fundamental quantity for connecting the observed pulsar population to the intrinsic Galactic population and for constraining pulsar emission geometry. In this study, we estimate the beaming fraction in each observational band (radio, $\gamma$-ray, and X-ray) and for each TeV survey (H.E.S.S., HAWC, and LHAASO) using TeV-selected pulsar wind nebulae (PWNe) and unidentified (Unid) TeV sources, assuming that the TeV emission from PWNe is approximately isotropic and that Unid sources are PWNe powered by pulsars whose beams do not intersect our line of sight. Within each survey, the inferred beaming fractions $\sim 0.1-0.3$ are comparable across bands. In contrast, the values differ by more than a factor of two between H.E.S.S. and HAWC/LHAASO. This discrepancy likely reflects survey-dependent selection effects, including differences in angular resolution and energy range, and is also consistent with the possibility that HAWC/LHAASO selected samples preferentially include older pulsars associated with more extended PWNe than those in the H.E.S.S. sample. We further show that the inferred beaming fractions can be reproduced within a unified framework using a time-dependent opening angle, and that this framework remains compatible with the statistical properties of the observed pulsar population.
\end{abstract}



\section{Introduction}\label{sec:Intro}

Pulsars are rotation-powered neutron stars, typically observed with radio, and currently, the number
of observed radio pulsars is roughly 4000\footnote{\href{https://www.atnf.csiro.au/research/pulsar/psrcat/}{https://www.atnf.csiro.au/research/pulsar/psrcat/}} \citep{Manchester2005}. On the other hand, pulsars are also bright bright in X-rays, optical, and $\gamma$-rays \citep[for a review, see][]{Enoto2019}. \textit{Fermi} Large Area Telescope (LAT) has revealed more than 300 $\gamma$-ray pulsars, including radio-quiet and millisecond pulsars \citep{Abdo2013(2ndFermi),Smith2019,Smith2023(3rdFermi)}. Despite these advances, the origin and geometry of pulsar emission remain poorly understood.

Radio emission from pulsars is phenomenologically believed to arise from a polar cap region \citep[][]{Sturrock1971,Ruderman1975}.
The polar cap cascade is thought to produce a relativistic plasma outflow along the open magnetic field lines, naturally leading to a narrow radio beam \citep[][]{Weltevrede2008, Johnston2019, Mitra2023}.

In the case of $\gamma$-rays pulsar, based on the spectral properties of the gamma-ray emission, it is likely that the radiation does not originate from the polar cap \citep{Abdo2013(2ndFermi)}, but rather from the outer region such as the slot gap \citep[formed along the edges of the open magnetic field lines;][]{Arons1983,Muslimov2003,Muslimov2004} , the outer gap \citep[around light cylinder;][]{Cheng1986, Takata2004}, and the striped wind \citep[][]{Kirk2002, Bai2010}. 

An important quantity that bridges theoretical beam geometry models and observations is the beaming fraction \citep[][]{Lyne1988,Tauris1998,Zhang2003,Kolonko2004,Ravi2010,Watters2009,Johnston2020,Turner2025}. 
The beaming fraction is defined as the fraction of the full solid angle swept by the pulsar’s radio beam during one rotation, and it quantifies the probability that a randomly oriented observer’s line of sight will intersect the beam. 
The beaming fraction serves as a key correction factor for reconstructing the observed radio and $\gamma$-ray pulsar populations to the true Galactic population in Monte-Carlo simulation \citep[MC; e.g.][]{Faucher2006}. 

The assumptions about the beaming fraction are also critical in estimating the double neutron star (DNS) merger rate and the gravitational-wave detection rate \citep[][]{Kim2003,Shaughnessy2010,Oslowski2011,Pol2019,Pol2020,Grunthal2021}. 
By means of population synthesis models that adopt the radio pulsar beaming fraction \citep{Tauris1998}, latest studies \citep{Grunthal2021} predict the expected detection rate of DNS mergers by the LIGO to be $R_{\mathrm{LIGO}} =4.6^{+7.1}_{-3.4}\,D_{r,100\,\mathrm{Mpc}}^3~\mathrm{yr^{-1}} $,
where $D_r$ is the luminosity distance. 
However, this prediction stands in stark contrast to the observational reality:
despite more than three years of continuous operation during the O4 run,
no confirmed gravitational-wave event associated with a DNS merger has yet been detected so far \citep[e.g.][]{LIGO2025}.
This discrepancy suggests that one or more systematic uncertainties, including the radio beaming fraction, may be larger than previously thought \citep{Pol2019, Grunthal2021,Fishbach2026}. 

Previous studies have estimated the radio or $\gamma$-ray pulsar beaming fractions by mean of  various methods. 
\citet{Lyne1988} and some studies \citep{Tauris1998, Zhang2003, Kolonko2004} modeled the opening angle as a function of spin period and obtained an integrated beaming fraction of $f_b \sim 0.2-0.3$ for radio pulsars \citep{Lyne1988, Zhang2003, Kolonko2004} or beaming fraction as a function of spin-period and spin-down age \citep{Tauris1998}.
\citet{Ravi2010} adopted a different statistical approach by mean of  the population of $\gamma$-ray pulsars detected by \textit{Fermi}-LAT. 
Assuming that $\gamma$-ray emission is observable regardless of viewing geometry, they compared the numbers of radio-loud and radio-quiet pulsars above given thresholds in spin-down luminosity $\dot{E}_{\mathrm{th}}$, 
and derived cumulative beaming fractions with a maximum of $f_b \sim 0.5$. Note that \citet{Ravi2010} is restricted to the population of $\gamma$-ray pulsars detected by \textit{Fermi}-LAT.
\citet{Watters2009} derived an analytic expression for the beaming fraction of $\gamma$-ray pulsars by assuming geometric beam models, including outer-gap and two-pole caustic models. 
They estimated the magnetic inclination angle and viewing angle distributions for both radio-loud and radio-quiet populations and used their relationship with the beam opening angle to quantify the analytical beaming fraction as a function of magnetic inclination angle and viewing angle across different beam models.
\citet{Johnston2020} adopts MC simulations to estimate the beaming fractions of both radio and $\gamma$-ray pulsars as functions of spin-down luminosity, 
assuming the polar-cap model for radio pulsars and the outer-gap model for $\gamma$-ray pulsars, together with spin-period evolution. 
Their results are in line with the ratio of radio-loud to $\gamma$-ray pulsars derived by \citet{Ravi2010}. 
Most studies have inferred the beaming fraction from observed radio and $\gamma$-ray pulsars by adopting specific beam geometry models.
Consequently, the resulting values are highly model-dependent theoretical estimates and may not accurately represent the intrinsic emission geometry shared by the whole pulsar population. Even when restricted to young pulsars, estimating of MC simulation remain sensitive to the assumed beam model, luminosity function, and inclination angle evolution; accordingly, both the methodology and the interpretation of the estimated beaming fraction require caution.

To mitigate these model-dependent limitations, we focus on pulsar wind nebulae (PWNe) which are bubbles of relativistic particles powered by pulsar spin-down and co-evolving with their supernova remnants \citep[SNR; e.g.,][]{Bucciantini2011, Gaenslar2006, Kargaltsev2015}.
In reality, pulsar winds are intrinsically anisotropic, often concentrated near the equatorial plane in the form of tori or rings \citep[][]{Kaspi2001} and accompanied by polar jets \citep{Komissarov2004, Kargaltsev2007}. The degree of anisotropy depends on the orientation of the pulsar’s spin axis and toroidal magnetic field relative to its kick velocity, which can give rise to elongated or asymmetric nebular morphologies \citep[][]{Begelman1992, Fesen1992, Olmi2016}.
Nevertheless, at TeV energies, PWNe radiate primarily via inverse-Compton emission from relativistic electrons distributed throughout the nebula. Since the emitting region 
is effectively optically thin and non-relativistic bulk velocity, the TeV emission is expected to be isotropic. 
Therefore, PWNe provide an indirect tracer estimating the beaming fraction and exploring the pulsar population.

In line with our method, \citet{Turner2025} recently estimated the integrated value of $f_b \sim 0.1\,-\,0.3$ using SNR associations by comparing the numbers of associated and unassociated pulsars based on the results of the TRAPUM \citep[TRAnsients And PUlsars with MeerKAT;][]{Turner2024, Turner2025} search. They further tested its consistency with the beaming function derived from \citet{Tauris1998} via MC simulations and found broad agreement. Our PWN-based approach is conceptually similar to the SNR-based approach; these do not require explicit assumptions about the emission geometry. However, the detectability of SNRs is strongly influenced by factors such as distance, line-of-sight absorption, contamination near the Galactic plane, ambient density, angular size, and surface-brightness evolution.
In the case of PWNe, the sample is not contaminated by Type Ia supernovae, and PWNe are generally easier to identify because they are bright at TeV energies and are expected to emit nearly isotropic, unlike many shell-like SNRs that are often difficult to identify morphologically.

This paper is organized as follows. 
In Section~\ref{sec:Method}, we describe the methodology used to estimate the pulsar beaming fraction based on the statistical comparison between PWNe and unidentified TeV sources. 
In Section~\ref{sec:result}, we present the derived beaming fractions for each survey, as well as their dependence on pulsar parameters such as spin-down luminosity and characteristic age. 
We discuss our results in Section \ref{sec:discussion} and give conclusions in Section \ref{sec:conclusion}.


\section{Method}\label{sec:Method}
To estimate the Galactic pulsar beaming fraction, we construct two complementary samples based on \citet{Tevcat2008}: TeV sources identified as PWNe and unidentified (Unid) sources. The samples are based on TeV selected because the majority of Galactic TeV $\gamma$-ray sources are associated with PWNe, and only a limited number of other source classes contribute significantly in the TeV band \citep{HESScollaboration2018, HESS2018PWN, Kargaltsev2017}. Although AGNs are also known TeV emitters \citep{Dermer2016}, they are predominantly found at high Galactic latitudes. Therefore, restricting the sample to sources near the Galactic plane helps minimize contamination. The PWNe represent that  the pulsar beam intersects our line of sight, whereas the Unid sources are interpreted as potential pulsars whose beams do not intersect the observer’s line of sight.
In this study, we assume that TeV emission from PWNe is isotropic and independent of the pulsar's beaming geometry. Since the missing pulsars in Unid sources lack information on characteristic age and spin-down luminosity, we estimate the cumulative beaming fraction as the simple ratio of the number of Unid sources to PWNe.
We classify the PWNe by the observed band of the associated pulsars including radio, $\gamma$-ray and non-thermal X-ray based on \citep{Manchester2005}. Finally, we derive the beaming fraction independently for each survey, including H.E.S.S, HAWC, and LHAASO.

\begin{figure*}
    \includegraphics[width=2\columnwidth]{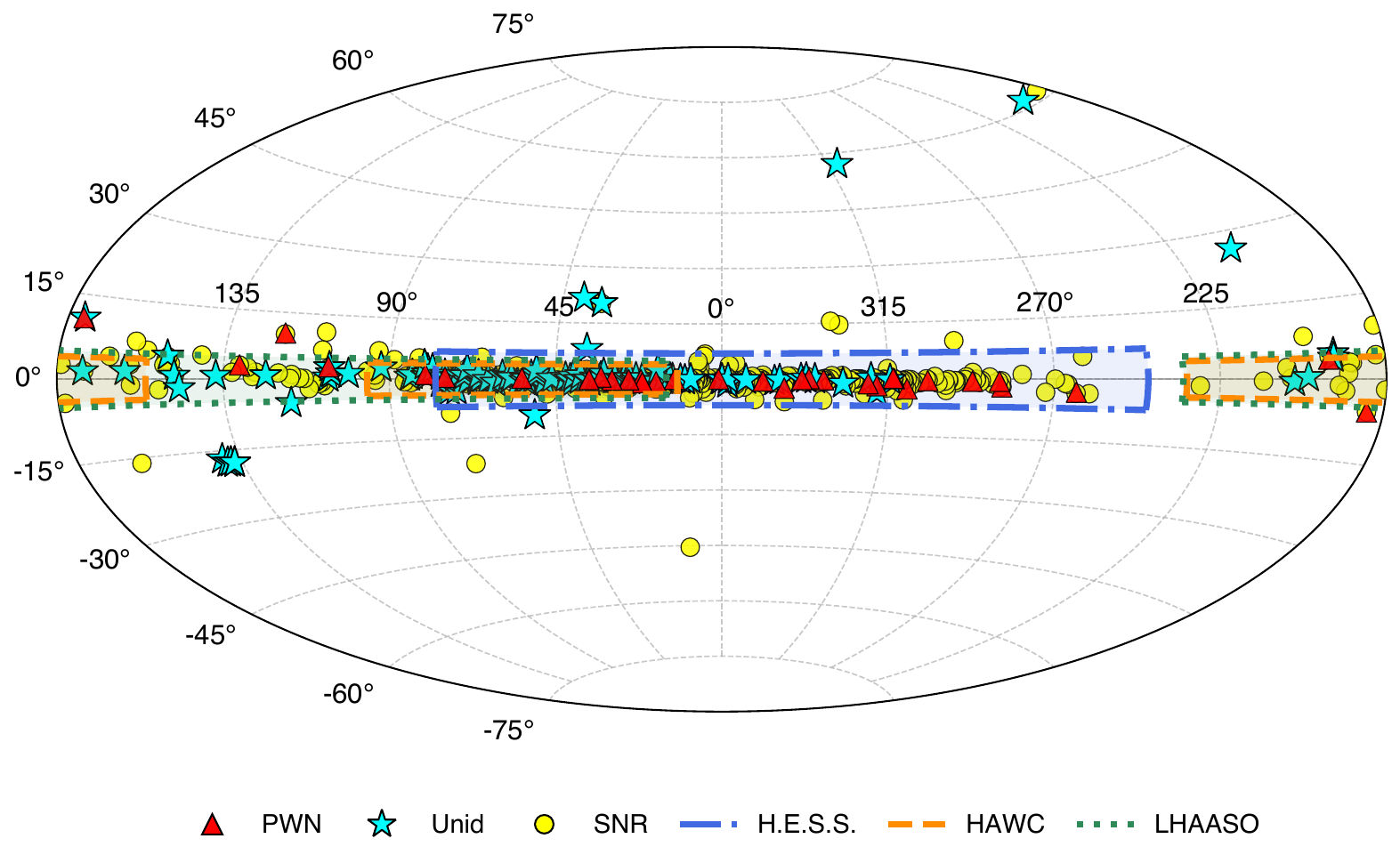
   }\caption{Galactic distribution of PWNe (triangle), Unid sources (star) and SNRs (circle) in Galactic coordinates. The outlines of the respective survey regions are overlaid for reference.
}
    \label{fig:mollweide}
\end{figure*}

\subsection{TeV-Selected PWNe }\label{subsec:TeV-selected PWNes}

We list the identified PWNe from TeVCat ver2.2 \footnote{\href{https://tevcat2.tevcat.org/}{https://tevcat2.tevcat.org/}} \citep{Tevcat2008} and comprises 38 entries in total. This list includes both firmly established PWNe and candidate sources that have been suggested based on their TeV properties. For the purpose of this study, we treat all of them uniformly as PWNe. Three sources, 
J1554-550, J1953+294, and J2031+415 
were excluded from PWNe samples. 
J2031+415 is considered to be associated with the $\gamma$-ray binary PSR J2032+415/MT91 213 \citep{Aharonian2005} and is known to overlap with another PWN candidate, J2032+415 \citep{Alfaro2024}.

While 
J1554-550 is interpreted as a PWN associated with SNR G327.1-1.1, whose pulsar has not been detected in any band \citep{Eagle2022} and it is likely associated with a pulsar whose beam does not intersect our line of sight.  J1953+294 associated with the SNR G65.7+1.2 is also in the same situation \citep{Coerver2019}. We therefore include J1554–550 \& J1953+294 among the Unid sources discussed in Section \ref{subsec:unid} and \ref{subsec:estimating}.

\begin{table*}
\centering
\captionsetup{width=\textwidth}
\caption{Physical and observational parameters of PWNe detailed in Sec.~\ref{subsec:surveys} and their associated pulsars, based on TeVCat \citep{Tevcat2008} for PWNe, the ATNF catalog \citep{Manchester2005} for radio pulsars, the \textit{Fermi}-LAT catalogs \citep{Abdo2013(2ndFermi),Smith2023(3rdFermi)} for \(\gamma\)-ray pulsars and some references for pulsars in X-ray.}
\label{tab:pwn_params}

\resizebox{\textwidth}{!}{%
\begin{tabular}{l l l l l c c c c c}
\hline
PWN name & Common name & ATNF name & Survey$\,^{\mathrm{(a)}}$ & band$\,^{\mathrm{(b)}}$ & $\log_{10}\dot{E}$$\,^{\mathrm{(c)}}$ & $\tau_{\mathrm{c}}$$\,^{\mathrm{(d)}}$ & $d$$\,^{\mathrm{(e)}}$ & $L_{1-10\,\mathrm{TeV}}$$\,^{\mathrm{(f)}}$ & refs.$\,^{\mathrm{(g)}}$ \\
\hline
J0632+173 & Geminga & J0633+1746 & Milagro, HAWC, LHAASO & radio, $\gamma$-ray, X-ray(pulsed) & 34.51 & 342.0 & 0.19 & $0.02 \pm 0.01$ & [1] \\
J0835-455 & Vela X & B0833-45 & H.E.S.S. & radio, $\gamma$-ray, X-ray(pulsed) & 36.84 & 11.3 & 0.28 & $0.58 \pm 0.02$ & [2],[3] \\
J1018-589B & The Goose & J1016-5857 & H.E.S.S. & radio, $\gamma$-ray, X-ray & 36.41 & 21.0 & 8.00 & $19.02 \pm 2.29$ & [4] \\
J1026-582 & -- & J1028-5819 & H.E.S.S. & radio, $\gamma$-ray, X-ray & 35.92 & 90.0 & 2.30 & $1.58 \pm 0.21$ & [5] \\
J1119-614 & G292.2$-$0.5 & J1119-6127 & H.E.S.S. & radio, $\gamma$-ray, X-ray(pulsed) & 36.36 & 1.61 & 8.40 & $24.38 \pm 2.95$ & [6] \\
J1303-631 & -- & J1301-6305 & H.E.S.S. & radio, X-ray & 36.23 & 11.0 & 6.65 & $89.94 \pm 2.30$ & [7] \\
J1356-645 & -- & J1357-6429 & H.E.S.S. & radio, $\gamma$-ray, X-ray(pulsed) & 36.49 & 7.31 & 2.50 & $11.63 \pm 1.02$ & [8] \\
J1418-609 & Kookaburra (Rabbit) & J1418-6058 & H.E.S.S. & $\gamma$-ray, X-ray(pulsed) & 36.69 & 10.3 & 5.00 & $28.26 \pm 1.69$ & [9],[10] \\
J1420-607 & Kookaburra (K2) & J1420-6048 & H.E.S.S. & radio, $\gamma$-ray, X-ray(pulsed) & 37.00 & 13.0 & 5.61 & $37.11 \pm 2.08$ & [11] \\
J1459-608 & -- & J1459-6053 & H.E.S.S. & $\gamma$-ray, X-ray(pulsed) & 35.96 & 64.7 & 1.84 & $2.14 \pm 0.52$ & [7],[12] \\
J1514-591 & MSH 15-52 & B1509-58 & H.E.S.S. & radio, $\gamma$-ray, X-ray(pulsed) & 37.23 & 1.56 & 4.40 & $44.02 \pm 1.63$ & [13] \\
J1616+508 & RCW 103 & J1617$-$5055 & H.E.S.S. & radio, X-ray(pulsed) & 37.20 & 8.13 & 4.74 & $209.77 \pm 16.25$ & [14] \\
J1632-478 & -- & J1632-4757 & H.E.S.S. & radio & 36.48 & 20.0 & 3.00 & $8.12 \pm 0.63$ & [15] \\
J1640-465 & -- & J1640-4631 & H.E.S.S. & X-ray(pulsed) & 36.64 & 3.35 & 12.8 & $162.01 \pm 8.25$ & [16] \\
J1708-443 & -- & B1706-44 & CANGAROO, H.E.S.S. & radio, $\gamma$-ray, X-ray(pulsed) & 36.53 & 17.5 & 2.60 & $9.61 \pm 1.11$ & [17] \\
J1718-385 & -- & J1718-3825 & H.E.S.S. & radio, $\gamma$-ray, X-ray & 36.10 & 90.0 & 3.48 & $4.84 \pm 0.49$ & [18] \\
J1747-281 & G000.9+00.1 & J1747-2809 & H.E.S.S., VERITAS & radio, X-ray(pulsed) & 37.63 & 5.3 & 8.10 & $17.46 \pm 3.33$ & [19] \\
J1813-178 & G21.5$-$0.9 & J1813-1749 & HAWC, MAGIC, H.E.S.S. & radio, X-ray(pulsed) & 37.75 & 5.6 & 6.15 & $18.07 \pm 1.24$ & [20] \\
J1825-137 & -- & B1823-13 & HAWC, VERITAS, LHAASO, H.E.S.S. & radio & 36.45 & 21.4 & 3.93 & $111.26 \pm 4.64$ & [21] \\
J1831-099 & -- & J1831-0952 & HAWC, LHAASO, H.E.S.S. & radio, $\gamma$-ray, X-ray & 36.04 & 128.0 & 3.68 & $14.07 \pm 1.91$ & [22] \\
J1833-105 & MSH 15-52 & J1833-1034 & H.E.S.S. & radio, $\gamma$-ray, X-ray & 37.53 & 4.85 & 4.10 & $1.77 \pm 0.46$ & [23] \\
J1837-0698 & -- & J1838-0655 & LHAASO, HAWC, H.E.S.S. & X-ray & 36.74 & 22.7 & 6.60 & $193.20 \pm 10.00$ & [24] \\
J1846-029 & Kes 75 & J1846-0258 & H.E.S.S. & $\gamma$-ray, X-ray(pulsed) & 36.91 & 0.73 & 5.80 & $6.44 \pm 0.81$ & [7],[25] \\
J1849-000 & IGR J18490-0000 & J1849-0001 & HAWC, LHAASO, H.E.S.S. & X-ray(pulsed) & 36.99 & 42.9 & 7.00 & $12.49 \pm 1.56$ & [26] \\
J1857+026 & -- & J1856+0245 & H.E.S.S., MAGIC, HAWC, LHAASO & radio, X-ray & 36.66 & 20.6 & 6.32 & $160.96 \pm 69.31$ & [27] \\
J1930+188 & -- & J1930+1852 & H.E.S.S. & radio, X-ray(pulsed) & 37.08 & 2.89 & 7.00 & $5.98 \pm 1.40$ & [28] \\
J2019+368 & -- & J2021+3651 & Milagro, VERITAS, LHAASO & radio, $\gamma$-ray, X-ray(pulsed) & 36.53 & 17.2 & 1.80 & $10.01 \pm 2.86$ & [29],[30] \\
J2032+415 & -- & J2032+4127 & Crimea, Milagro, HAWC, MAGIC, LHAASO, etc. & radio, $\gamma$-ray, X-ray & 35.18 & 200.0 & 1.33 & $3.83 \pm 0.47$ & [31],[32] \\
\hline
\end{tabular}%
}

\vspace{2pt}
\begin{minipage}{0.95\textwidth}
\footnotesize
${\mathrm{(a)}}$ Survey(s) in which the PWN is detected.\\
${\mathrm{(b)}}$ Energy band(s) of the detected pulsar. For pulsars detected in non-thermal X-rays $> 2\,\mathrm{keV}$, we note ``pulsed'' if pulsations have been detected.\\
${\mathrm{(c)}}$ Spin-down luminosity (erg\,s\(^{-1}\)).\\
${\mathrm{(d)}}$ Characteristic age (kyr).\\
${\mathrm{(e)}}$ Pulsar distance (kpc).\\
${\mathrm{(f)}}$ The 1--10\,TeV luminosity given with \(1\sigma\) uncertainty in units of \(10^{33}\,\mathrm{erg\,s^{-1}}\).\\
${\mathrm{(g)}}$ References for the X-ray counterparts or pulsations:
[1] \citet{Halpern1992};
[2] \citet{Pavlov2001};
[3] \citet{Sanwal2002};
[4] \citet{Klingler2022};
[5] \citet{Mignani2012};
[6] \citet{Gonzalez2005};
[7] \citet{Kupier2015};
[8] \citet{Esposito2007};
[9] \citet{Kim2020};
[10] \citet{Park2023_1418};
[11] \citet{Park2023_1420};
[12] \citet{Pancrazi2012};
[13] \citet{Livingstone2011};
[14] \citet{Becker2002};
[15] \citet{Prinz2015};
[16] \citet{Gotthelf2014};
[17] \citet{Gotthelf2002};
[18] \citet{Hinton2007};
[19] \citet{Brunelli2026};
[20] \citet{Gotthelf2009};
[21] \citet{Gaensler2003};
[22] \citet{Vahdat2022};
[23] \citet{Camilo2006};
[24] \citet{Gotthelf2008};
[25] \citet{Kuiper2018};
[26] \citet{Gotthelf2011};
[27] \citet{Hessels2008};
[28] \citet{Camilo2002};
[29] \citet{Abdo2009};
[30] \citet{Li2018};
[31] \citet{Camilo2009};
[32] \citet{Li2017}.
\end{minipage}
\end{table*}

\subsection{ TeV-selected Unidentified Sources }\label{subsec:unid}

We further list 126 Unid sources from TeVCat. Although this sample likely contains a mixture of source classes, including SNR shells, X-ray binaries, and AGN. We assume that all Unid sources are PWNe powered by pulsars whose emission beams do not intersect with our line of sight. This assumption corresponds to an upper limit on the number of PWNe in the current sample, simultaneously a lower limit on the beaming fraction as we see in Sec. \ref{subsec:estimating}.
As discussed in Section~\ref{subsec:TeV-selected PWNes}, J1554$-$550 and J1953+294 are PWNe associated with SNRs but without pulsars, and we treat them as Unid sources.




\subsection{Estimating the pulsar beaming fraction}\label{subsec:estimating}
In this study, we assume that the TeV emission from PWNe is isotropic, such that the observed PWN is independent of both the beaming geometry of the pulsar and the intrinsic geometry of the nebula. In addition, we assume that the PWNe associated with pulsars whose emission beams intersect our line of sight, whereas the Unid sources are interpreted as potential PWNe associated with pulsars whose beams do not intersect the observer.
Under this assumption, the ratio of the number of PWNe to the total number of TeV sources can be regarded as the pulsar beaming fraction.


To investigate the band dependency of pulsar emission geometry, we classify the PWNe based on the observed band of associated pulsar (radio, $\gamma$-ray, or X-ray). For the radio and \(\gamma\)-ray bands, we consider only pulsars with detected pulsations, based on the ATNF catalog \citep{Manchester2005} for radio pulsars and the \textit{Fermi}-LAT catalogs \citep{Abdo2013(2ndFermi),Smith2023(3rdFermi)} for \(\gamma\)-ray pulsars. The pulsars in X-ray  include both pulsars with detected X-ray pulsations and those without detected X-ray pulsations, as summarized in Table~\ref{tab:pwn_params}.

We define the pulsar beaming fraction, $f_{b,\,\nu}$, as
\begin{align}
\label{eq:beaming_fraction}
f_{b,\nu} = \frac{N_{\nu,\,\mathrm{PWN}}}{N_{\mathrm{PWN}} + N_{\mathrm{Unid}}},
\end{align}
where $N_{\mathrm{PWN}}$ is the number of TeV-selected PWNe, $N_{\nu,\,\mathrm{PWN}}$ is the number of TeV-selected PWNe associated with pulsars detected in a given band, and $N_{\mathrm{Unid}}$ is the number of Unid TeV sources. We note that the beaming fraction derived from Eq. (\ref{eq:beaming_fraction}) should be regarded as a lower limit for the current samples.
This is because $N_{\mathrm{PWN}}$ is unlikely to decrease with future observations,
whereas $N_{\mathrm{Unid}}$ may decrease as further analyses or higher-sensitivity observations classify them as PWNe or other TeV emitters.
Such reclassification would effectively reduce the denominator in Eq. \ref{eq:beaming_fraction}, leading to a larger $f_b$.

We also estimate the cumulative beaming fractions as a function of spin-down luminosity ($\dot{E}$) and characteristic age ($\tau_{\mathrm{c}}$).
The Unid sources lack measured values of  $\dot{E}$ and $\tau_{\mathrm{c}}$. Thus, we adopt for a given threshold that all Unid sources satisfies the threshold. 
We assume that Unid sources have spin-down luminosities above a given threshold ($\dot{E} > \dot{E}_{\mathrm{th}}$) and characteristic ages below a given threshold ($\tau_{\mathrm{c}} < \tau_{c,\mathrm{th}}$). 
Under this assumption, the cumulative beaming fraction is calculated as
\begin{align}
    \label{eq:cumulative_beaming_frction}f_{b,\nu, i_{\mathrm{th}}} = 
    \frac{N_{\nu,\,\mathrm{PWN},\, i \gtrless i_{\mathrm{th}}}}
         {N_{\mathrm{PWN},\, i \gtrless i_{\mathrm{th}}} + N_{\mathrm{Unid},\, i_{\mathrm{th}}}},
\end{align}
where $i$ denotes the pulsar parameter either spin-down luminosity (>) or characteristic age (<), and $i_{\mathrm{th}}$ is the corresponding threshold. We also note that the beaming fraction derived from Eq.~(\ref{eq:cumulative_beaming_frction}), when a threshold is imposed, should be regarded as a lower limit.

The statistical uncertainties on $f_b$ are estimated assuming Poisson noise in the number counts of both PWNe and Unid sources. 
These errors are propagated through Eq. (\ref{eq:beaming_fraction}) to derive the corresponding confidence intervals for each cumulative beaming fraction.

\subsection{Surveys}\label{subsec:surveys}

Here, we classify the PWN and Unid samples according to the TeV $\gamma$-ray surveys—H.E.S.S. \citep{HESScollaboration2018}, HAWC \citep{Abeysekara2017A(HAWC)}, and LHAASO \citep[][]{Cao2019(LHAASO)}—and estimate the beaming fraction for each survey independently. This approach is motivated by the substantial differences among the three surveys in sensitivity, angular resolution, exposure time, sky coverage, and energy band. Combining sources from different surveys without accounting for these differences would mix heterogeneous selection effects and could therefore bias the inferred beaming fraction.
By estimating the beaming fraction separately for each survey, we obtain a more internally consistent and physically interpretable comparison across the datasets.

Figure~\ref{fig:mollweide} shows the all-sky distribution of the PWNe and Unid sources \citep[][]{Tevcat2008}, with SNRs from SNRcat \citep[][]{Ferrand2012}.
The rectangular boxes indicate the Galactic-plane survey regions, while the blue dash-dotted, orange dashed, and green dotted lines represent the observational coverages of H.E.S.S., HAWC, and LHAASO, respectively.
The survey regions of H.E.S.S. are $ 0^\circ\leq l \leq 77.5^\circ$, $ 244.5^\circ\leq l \leq 360^\circ$ and $|b| \leq 7^\circ$ \citep{HESS2018PWN}.
The survey regions of HAWC are $ 12^\circ\leq l \leq 96^\circ$, $156^\circ \leq l \leq 234^\circ$ and $|b| \leq 4^\circ$ \citep{Albert2020(3HWC)}. The survey regions of LHAASO are
$ 15^\circ\leq l \leq 225^\circ$ and $|b| \leq 5^\circ$ \citep[][]{Cao2024}.

H.E.S.S.\ identifies TeV PWNe using four criteria \citep{HESS2018PWN}, (1) the pulsar offset must be $<1.5$ extension radii, (2) the source extension follows within $2\sigma$ of the size and age relation ($\sigma_{\ln R}=0.39$) (3) the TeV luminosity must be within $2\sigma$ of the TeV-luminosity and spin-down luminosity relation ($\sigma_{\ln L}=0.83$) (4) the TeV surface brightness must be within $2\sigma$ of the surface-brightness and spin-down luminosity relation ($\sigma_{\ln S}=0.30$) based on thier own model \citep[for a detail, see ][]{HESS2018PWN} . 
They classified sources satisfying all four criteria as PWNe, while sources satisfying at least one of them were classified as PWN candidates. However, many of these candidates are likely to be PWNe \citep{HESS2018PWN}, and TeVCat also classifies such candidates as PWNe \citep{Tevcat2008}. Motivated by this, we include PWN candidates in the PWN sample. 

As noted in Sec.~\ref{subsec:unid}, we assume that Unid sources are PWNe whose associated pulsar beams do not intersect our line of sight. Using the H.E.S.S. criteria for PWNe, we test the validity of this assumption by examining how many Unid sources satisfy at least one of these criteria.
We assume that the pulsars potentially associated with Unid sources are located at a distance of 5 kpc, have a spin-down age of $\tau_{\mathrm{c}} \leq 10\,\mathrm{kyr}$, and a spin-down luminosity of $\dot{E} \ge 10^{36}\,\mathrm{erg\,s^{-1}}$ based on the peak values of PWN associated pulsar distributions \citep{Manchester2005,Tevcat2008}.
 H.E.S.S. employs four criteria for identifying PWNe; however, the first criterion requires an established pulsar association and therefore can never be fulfilled by Unid sources. We thus focus on criteria 2 -- 4. Criterion~2 requires that the PWN radius inferred under the above assumptions, $R_{\mathrm{PWN}}$, be smaller than the upper envelope of the size -- age relation, $R_{\mathrm{PWN,\,upper}}(\tau_{\mathrm{c}} = 10\,\mathrm{kyr})$. Criterion~3 requires that the TeV luminosity estimated from the observed flux exceed $L_{1-10\,\mathrm{TeV},\,\mathrm{lower}}(\dot{E}=10^{36}\,\mathrm{erg\,s^{-1}})$ including $2\sigma$ in the same model. Criterion~4 requires that the surface brightness derived from the observed flux and angular extent exceed $S_{\mathrm{lower}}(\dot{E}=10^{36}\,\mathrm{erg\,s^{-1}})$ in the same model. Under these assumptions, we find that all 35 Unid sources in H.E.S.S., all 55 in HAWC, and 65 of the 71 in LHAASO satisfy at least one of criteria 2--4. These fractions support the interpretation of Unid sources as systems associated with missing pulsars.

HAWC classifies a source as a PWN only if it has been previously identified as a PWN in $\gamma$-ray surveys (e.g., H.E.S.S., VERITAS) and is positionally detected in the Galactic plane survey \citep{Albert2020(3HWC)}. LHAASO classify a source as a PWN candidate when a known pulsar lies within $0.5^\circ$ and the pulsar spin-down luminosity is $\dot{E}>10^{34}\,\mathrm{erg\,s^{-1}}$ \citep[][]{Cao2024}. As we see above, HAWC/LHAASO do not apply original detection criteria. As a result, PWNe detected by HAWC/LHAASO tend to be effectively limited by already reported PWNe by other instruments (e.g., H.E.S.S. and VERITAS). This tendency leads to a systematic bias estimating the beaming fraction. Therefore, we restrict our PWN and Unid samples to sources within each instrument’s Galactic-plane survey regions.

The final PWN samples used in this study are summarized in Table~\ref{tab:pwn_params}. H.E.S.S. samples contain 25 PWNe including candidates and 35 Unid sources.
HAWC samples contain 8 PWNe and 44 Unid sources, while LHAASO samples contain 8 PWNe and 60 Unid sources.
It is therefore evident that both HAWC/LHAASO contain a considerably larger fraction of Unid TeV sources compared with H.E.S.S.

\section{Result}\label{sec:result}

\begin{figure*}
    \centering
    \includegraphics[width=0.3\textwidth]
    {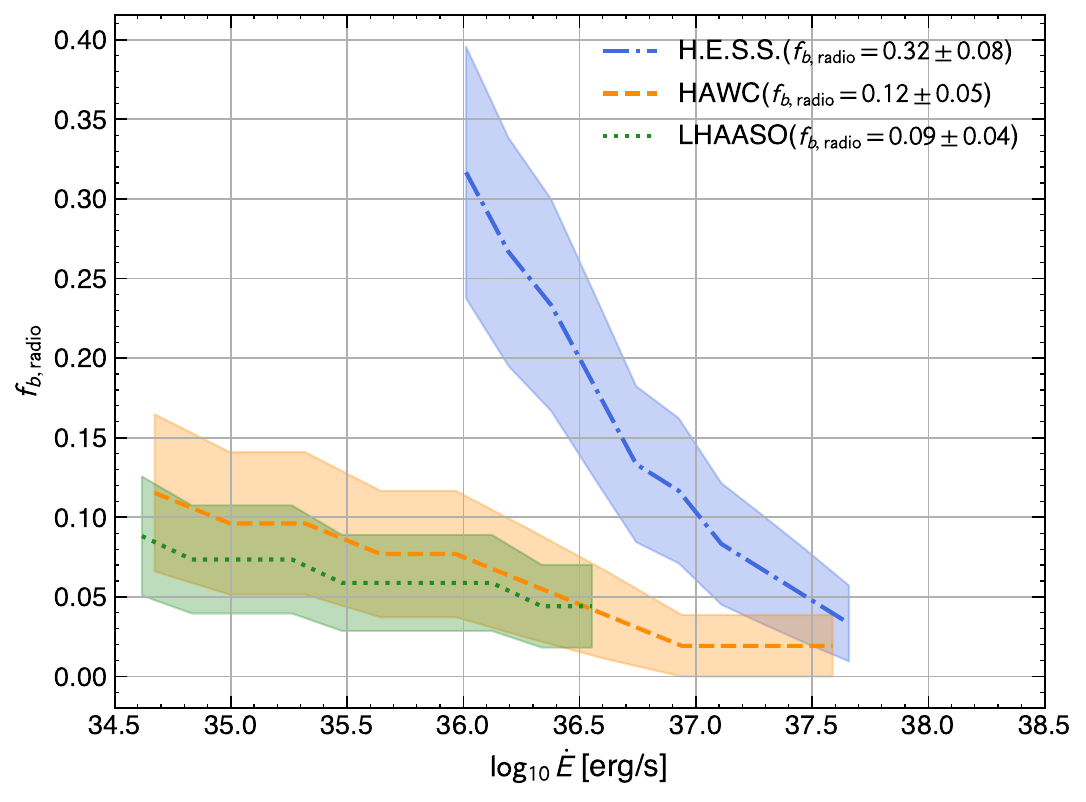}
    \includegraphics[width=0.3\textwidth]{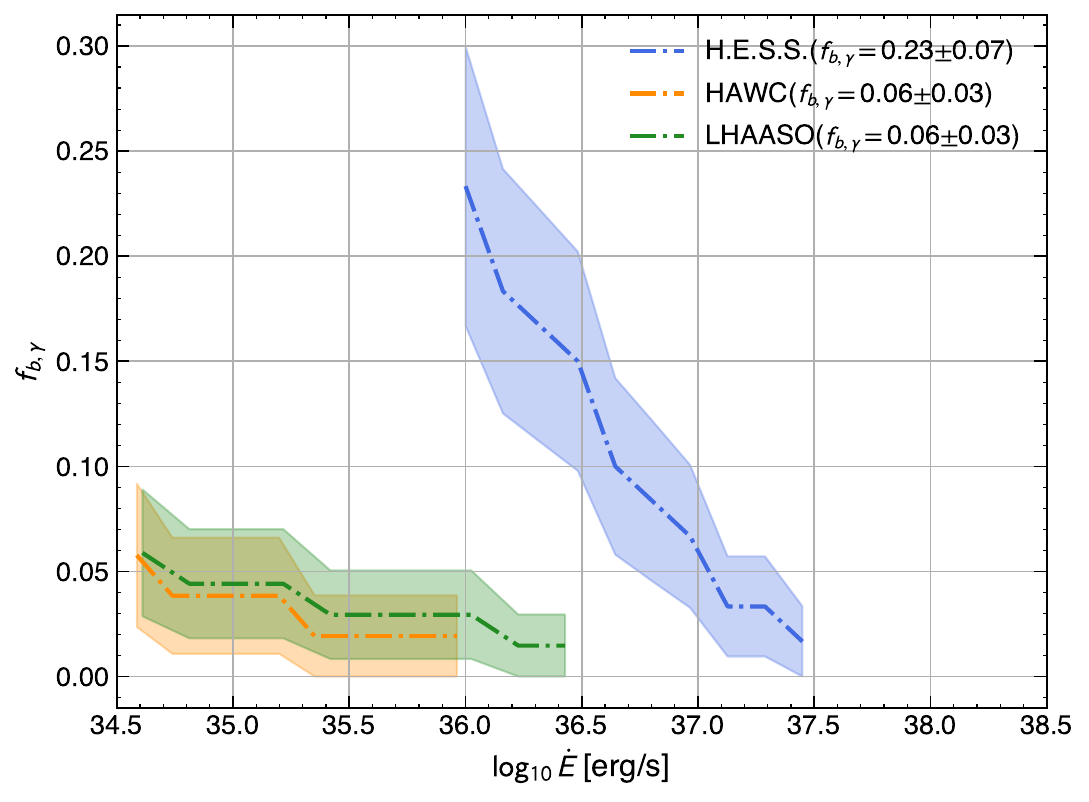}
    \includegraphics[width=0.3\textwidth]{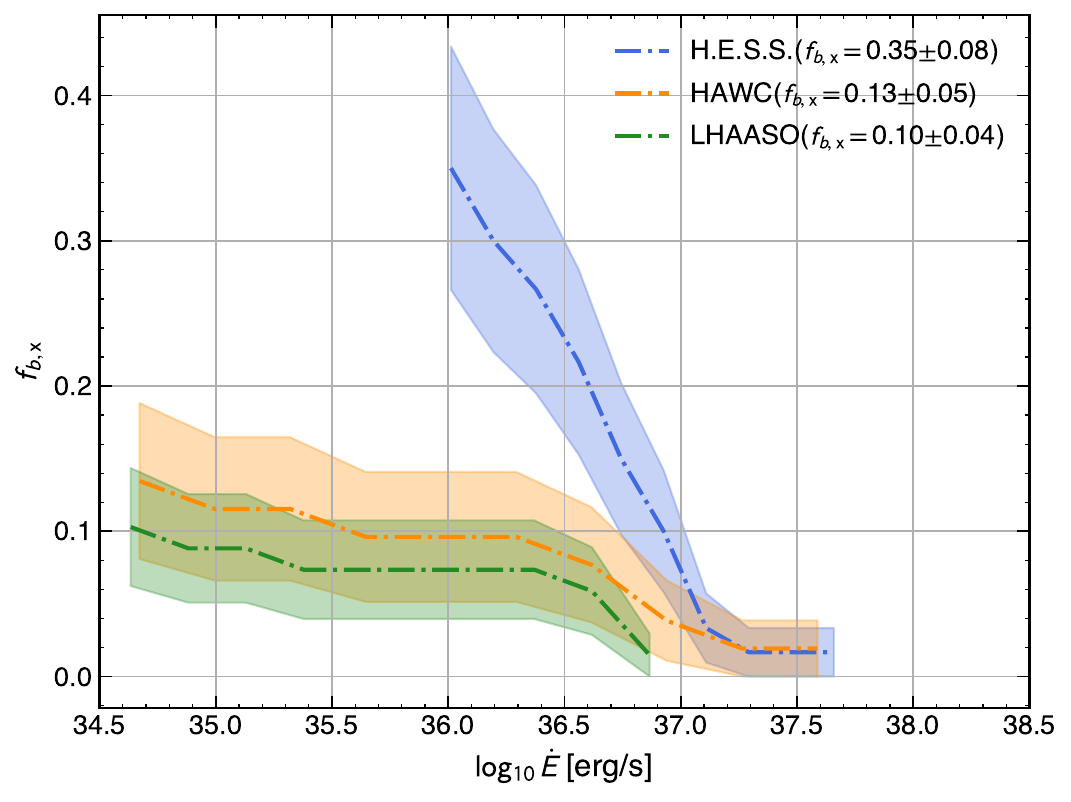}
    \caption{Cumulative beaming fractions as functions of the spin-down luminosity for pulsars observed at different bands, radio(left), $\gamma$-ray(center), and X-ray(right ; including pulsars without detected X-ray pulsations). The curves correspond to H.E.S.S. (dash-dotted), HAWC (dashed), and LHAASO (dotted), with the cumulative beaming fraction for all PWNe samples and Unid sources, $f_B$ indicated in the legend.}
    \label{fig:edot}
\end{figure*}

\begin{figure*}
    \centering
    \includegraphics[width=0.3\textwidth]
    {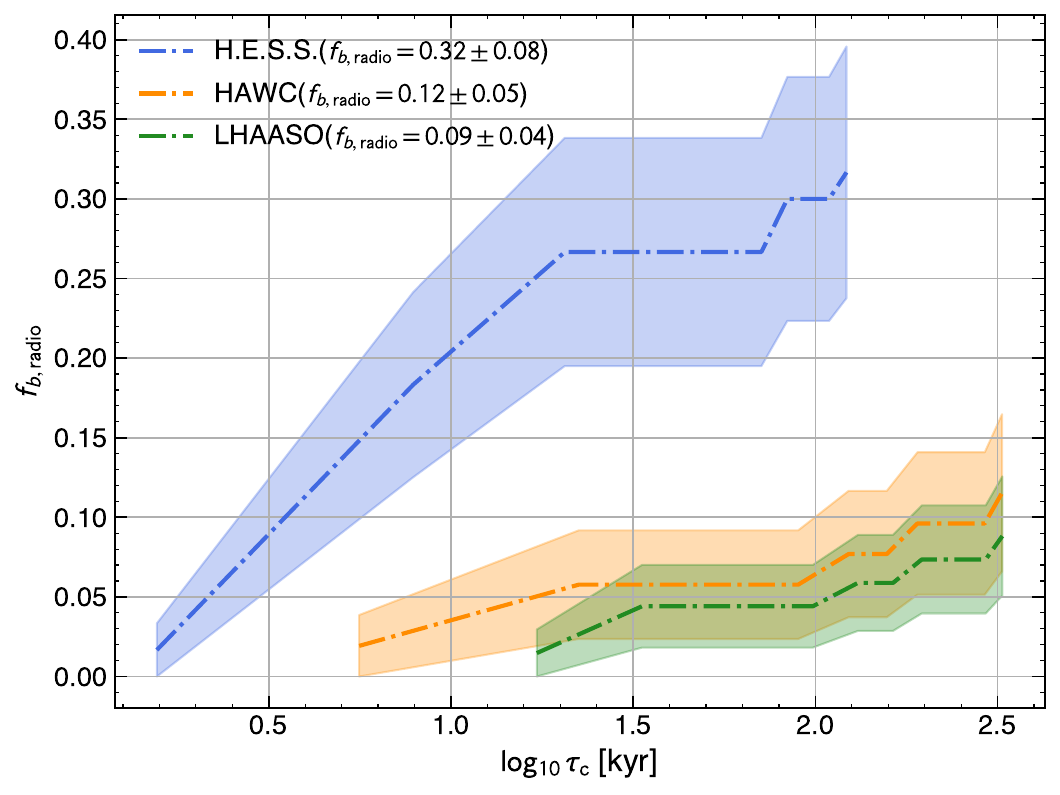}
    \includegraphics[width=0.3\textwidth]{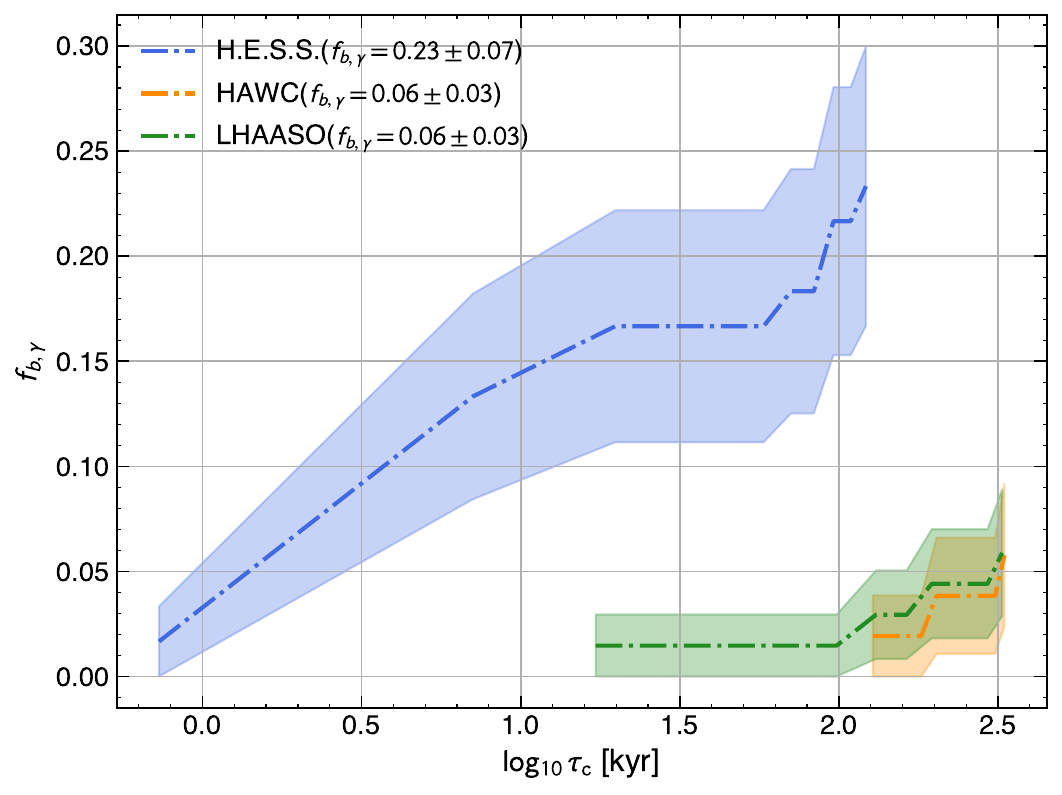}
    \includegraphics[width=0.3\textwidth]{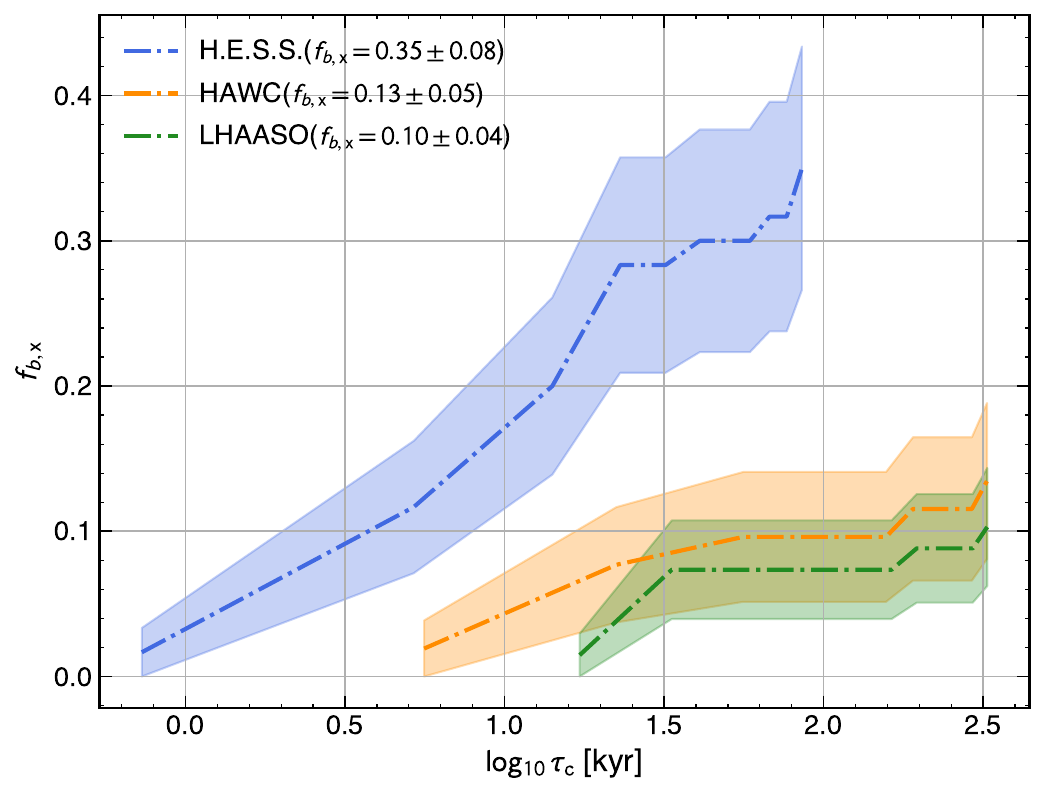}
    \caption{Similar to Fig. \ref{fig:edot}, but as a function of spin-down age.}
    \label{fig:age}
\end{figure*}

\begin{figure}
    \centering
    \includegraphics[width=0.65\linewidth]{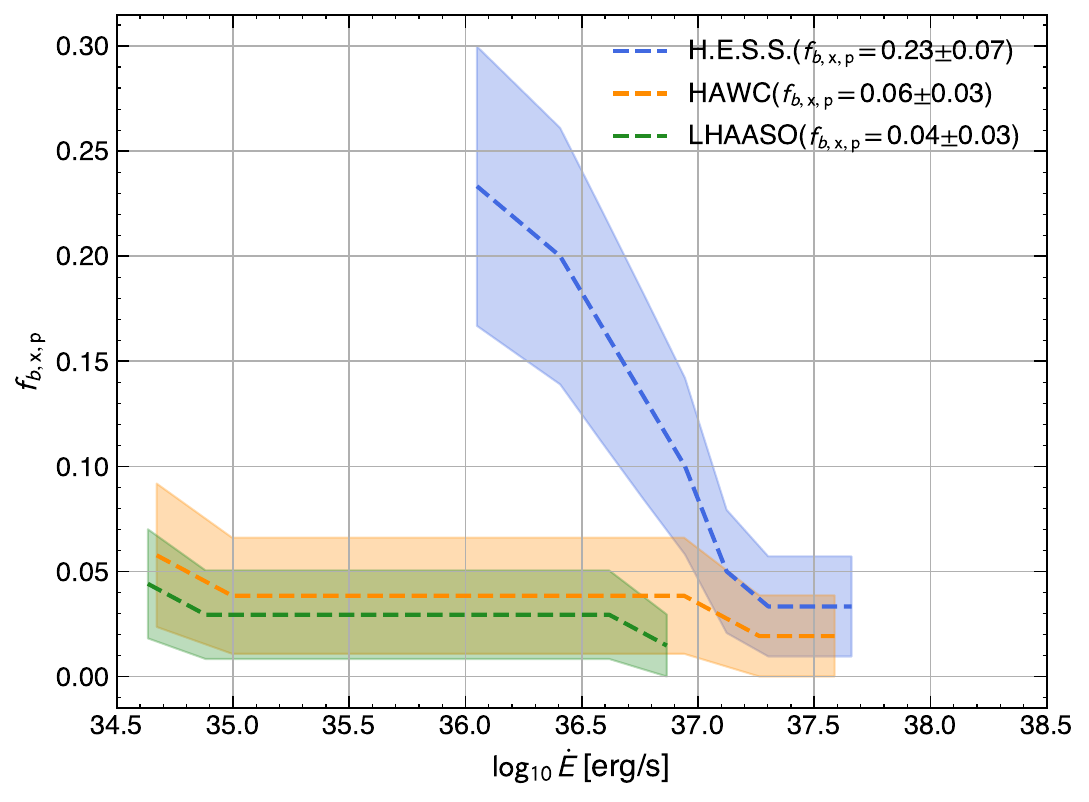}
    \includegraphics[width=0.64\linewidth]{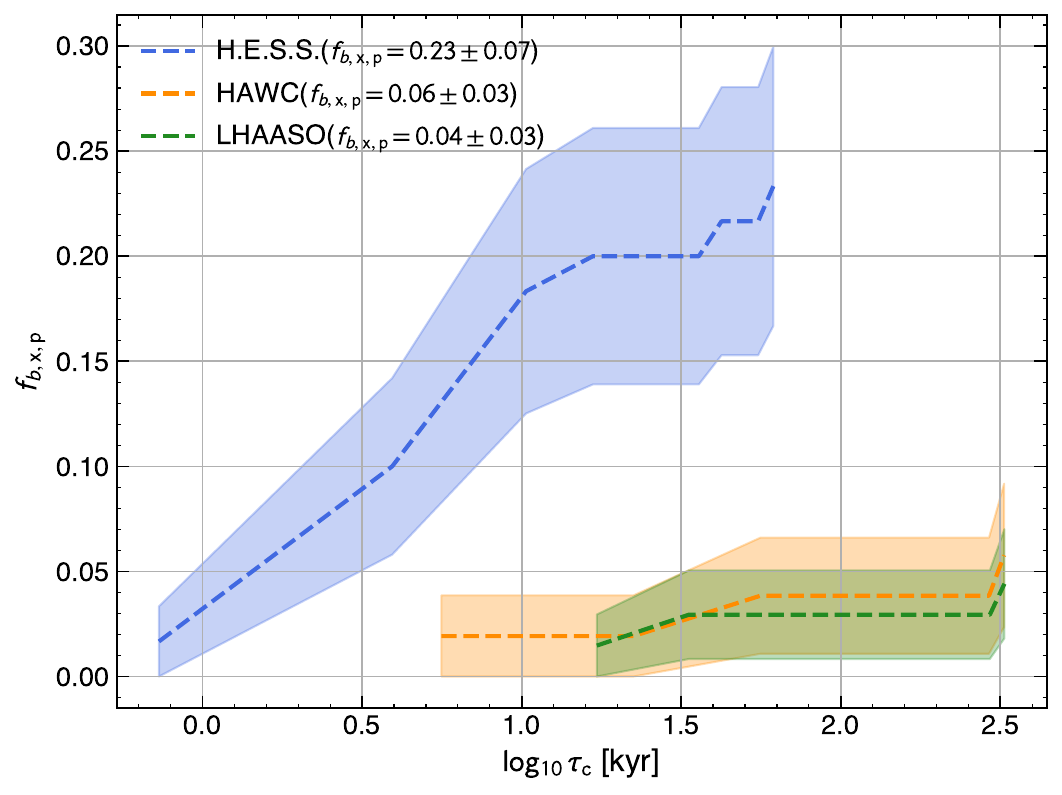}
    \caption{Similar to Fig. \ref{fig:edot} and \ref{fig:age}, but using only pulsars with detected X-ray pulsations.}
    \label{fig:pulse}
\end{figure}

\begin{table}
\centering
\caption{The beaming fraction of each surveys and bands based on Eq. (\ref{eq:beaming_fraction}).}
\label{tab:beam}
\begin{tabular}{ccccc}
\hline
Survey & radio & $\gamma$-ray & X-ray & X-ray(pulsed)\\
\hline
HESS & $0.32\pm0.08$ & $0.23\pm0.07$ & $0.35\pm0.08$ & $0.23\pm0.07$ \\ 
HAWC & $0.12\pm0.05$ & $0.06\pm 0.03$ & $0.13\pm0.05$ & $0.06\pm0.03$\\
LHAASO & $0.09\pm0.04$ & $0.06\pm 0.03$ & $0.10\pm 0.04$ & $0.04\pm0.03$  \\
\hline
\end{tabular}
\end{table}

We estimate the pulsar beaming fraction using TeV-selected PWNe and Unid sources. The beaming fractions obtained for each survey and each band are summarized in Table~\ref{tab:beam}. In addition, Figures~\ref{fig:edot} and \ref{fig:age} show the cumulative beaming fractions based on Eq.~(\ref{eq:cumulative_beaming_frction}), obtained using different physical quantities as thresholds.

Figure~\ref{fig:edot} shows the cumulative beaming fraction as a function of spin-down luminosity, under the assumption that Unid sources are PWNe associated with neutron stars whose spin-down luminosities exceed a threshold $\dot{E}_{\mathrm{th}}$. Figure~\ref{fig:age} shows the cumulative beaming fraction as a function of characteristic age, assuming that Unid sources are PWNe associated with neutron stars younger than a threshold $\tau_{c,\mathrm{th}}$. The error bars are based on Poisson noise.
In the case of H.E.S.S., the cumulative beaming fractions in the radio, $\gamma$-ray, and X-ray bands are mutually consistent within the error bars in the region where the spin-down luminosity is greater than $10^{36}\,\mathrm{erg/s}$. The same tendency is also seen in the HAWC/LHAASO samples.
On the other hand, the cumulative beaming fractions for HAWC/LHAASO extend to lower threshold values in spin-down luminosity and to older spin-down age. When a lower threshold in $\dot{E}_{\mathrm{th}} = 10^{34}\,\mathrm{erg/s}$ or an upper threshold in $\tau_{c,\mathrm{th}} =  10^{5.5}\,\mathrm{yr}$ is adopted, the resulting cumulative beaming fractions of HAWC/LHAASO are only about one-third to one-half of those inferred from H.E.S.S. Figure~\ref{fig:pulse} shows the same analysis as Figures~\ref{fig:edot} and \ref{fig:age}, but restricted to the sample of pulsars with detected X-ray pulsations. Compared with the case in which pulsars without detected X-ray pulsations are also included, the beaming fraction inferred from the sample detected pulsation with $\dot{E}_{\mathrm{th}} = 10^{34.5}\,\mathrm{erg/s}$ or $\tau_{\mathrm{c,\,\mathrm{th}}} = 10^{5.5}\,\mathrm{yr}$ is smaller by about a factor of two (also see Table. \ref{tab:beam}). This implies that a substantial fraction of X-ray pulsars does not have detected X-ray pulsations even when a non-thermal X-ray component is detected.

\section{Discussion}\label{sec:discussion}

\subsection{Instrumental Effects on the Derived Beaming Fractions}\label{subsec:Instrument}

A comparative analysis of the beaming fractions across different surveys indicates that the value obtained from H.E.S.S. is approximately 2–4 times larger than those obtained from HAWC/LHAASO.
Even after accounting for statistical uncertainties estimated from Poisson noise, this discrepancy remains statistically significant.
One possible explanation is instrumental effects.
In the following subsections, we enumerate and examine the potential factors that could explain the differences in the derived beaming fractions among the surveys.

(1) Angular resolution : In the case of H.E.S.S., the angular resolution (68\% containment radius of the point spread function) is approximately $0.048^{\circ}$ above 250 GeV and $0.036^{\circ}$  above 2 TeV, with a field of view of about $5^{\circ}$ \citep[][]{HESS2018PWN, HESS2018(RXJ1713.7-3946)}.
Then, some sources may appear extended beyond the field of view \citep{Bamba2010}.
In particular, PWNe expand as their SNRs evolve, and older systems with characteristic ages of $\geq 10$ kyr often reach exceeding $1^{\circ}$, becoming very extended sources that are more difficult to detect with H.E.S.S \citep[][]{HESS2018PWN}. 
In the case of HAWC, at energies between 300 GeV and 3 TeV, the 68\% containment radius of the point spread function is about $1^\circ$ at 3 TeV, improving to 0.2–$0.6^\circ$ above 10 TeV \citep[][]{Abeysekara2017(crab), Albert2020(3HWC), Alfaro2024}. 
The field of view covers approximately 15\% of the sky at any given time, allowing HAWC to survey about two-thirds of the entire sky every 24 hours \citep[][]{Albert2024(crab)}.
This wide-field design makes HAWC particularly sensitive to extended sources.
LHAASO also has a same advantage: its angular resolution is about 0.45° at 1 TeV and improves to 0.2° above 6 TeV, with a very large field of view of roughly 2 sr for the KM2A array \citep[][]{Cao2019(LHAASO), Aharonian2021, Cao2024}.
Indeed, LHAASO has already identified several PWNe with extensions $1^\circ$, demonstrating its capability for detecting very extended sources \citep[][]{Cao2024}. As a result, wide-field instruments such as HAWC/LHAASO are highly sensitive to nearby middle-aged pulsars associated with spatially extended TeV halos, whereas H.E.S.S. lose sensitivity to sources with large angular extents above $0.6^\circ$. This leads the two classes of instruments to selectively detect different TeV source populations, with HAWC being well suited to degree-scale halos such as Geminga and Monogem, which are the middle-aged systems \citep[][]{Linden2017}.
This tendency also reflects spin-down luminosity and characteristic age of samples in this study. In H.E.S.S., the cumulative distribution is primarily built up by young and high-$\dot{E}$ pulsars. By contrast, in HAWC/LHAASO, the cumulative distribution is built up by  old and low-$\dot{E}$ pulsars, indicating a broader sensitivity to evolved systems.

\begin{figure}
    \includegraphics[width=\linewidth]{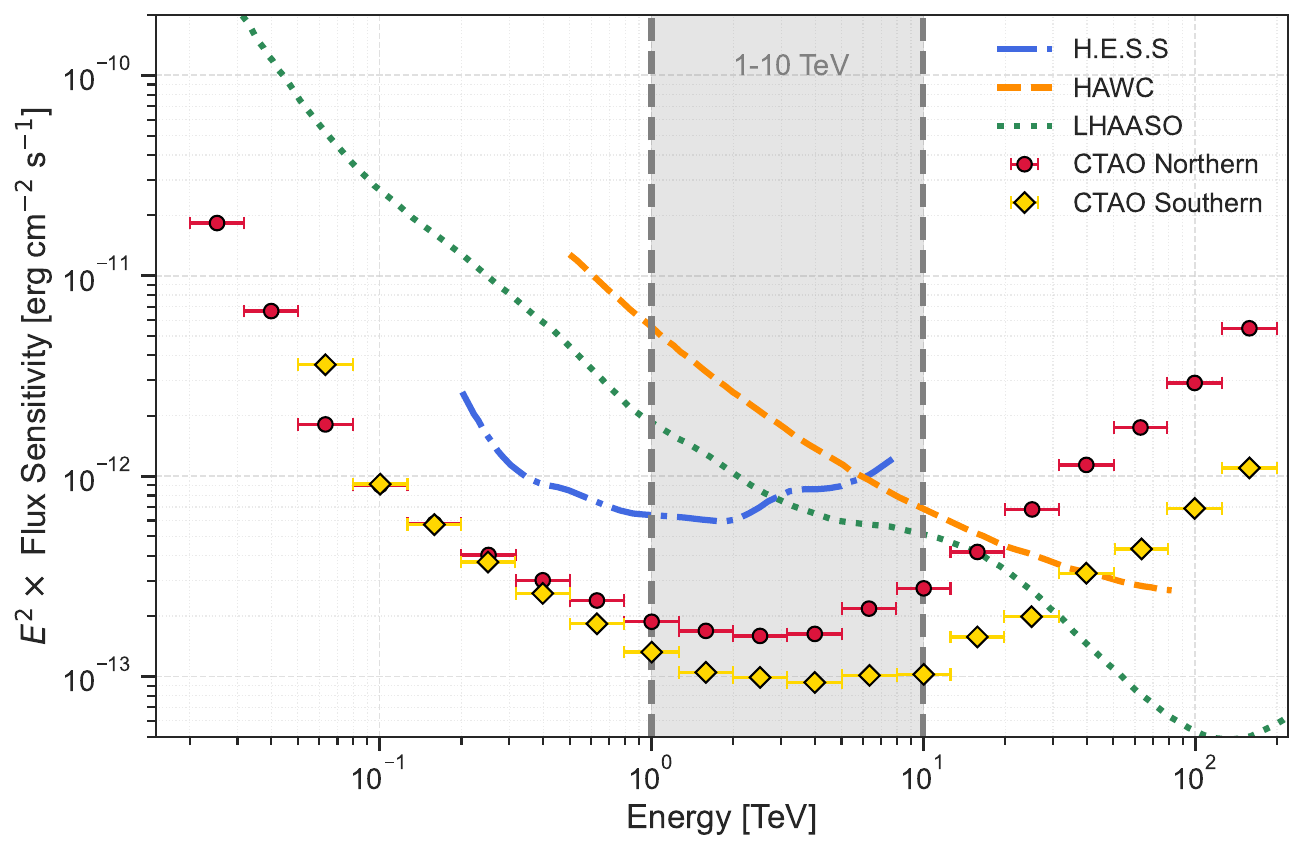}
    \caption{The differential sensitivities as a function of energy are shown for the following surveys: H.E.S.S. (50 h, dash-dotted line; \citealt{HESScollaboration2018}), HAWC (5 years, dashed line; \citealt{Alfaro2024(HAWC)}), LHAASO (1 year, dotted line; \citealt{Cao2019(LHAASO)}), and CTAO, with the Northern and Southern Arrays represented by circles and squares, respectively (50 h each; \citealt{CTAO2019}). We also indicate the 1–10 TeV range, which corresponds to the characteristic energy band of PWN spectra.}
    \label{fig:sensitivity}
\end{figure}

(2) Energy range : The flux sensitivity curves of H.E.S.S., HAWC, and LHAASO differ markedly in both energy coverage and limiting sensitivity. Figure \ref{fig:sensitivity} shows the differential sensitivities as a function of energy. H.E.S.S. achieves its highest sensitivity in the 1–10 TeV range but has a relatively narrow field of view and limited exposure time for individual sources.
In contrast, the wide-field air-shower arrays HAWC/LHAASO have lower sensitivity at lower energies, yet they achieve superior performance above 10 TeV and can continuously monitor a large fraction of the sky. 
HAWC/LHAASO have detected a large number of Unid sources, many of which exhibit spectral peaks above 10 TeV \citep{Abeysekara2017A(HAWC),Cao2019(LHAASO),Tevcat2008}. 
The difference in energy range leads to differences in the observed properties of PWNe. The TeV emission of PWNe is characterized by inverse Compton scattering. As PWNe evolve, the contribution from synchrotron self-Compton (SSC) decreases; since SSC primarily contributes in the lower energy of the inverse Compton component, the spectral peak shifts to higher energies with time \citep{Gelfand2009,Tanaka2010,Bucciantini2011}. 
H.E.S.S. is most sensitive in the 1--10 TeV range, and therefore preferentially detects relatively young PWNe whose spectral peaks lie at lower energies. In contrast, HAWC/LHAASO are more sensitive above 10 TeV, and thus may preferentially include more evolved PWN candidates whose spectral peaks have shifted to higher energies. As PWNe evolve, their surface brightness decreases due to expansion, making them more difficult to detect. Consequently, the TeV populations sampled by different observatories are not identical. 


(3) Contamination from background: In regions with high source density, such as the Galactic plane, flux contamination from nearby sources is a well-known systematic effect, particularly for instruments with large point-spread functions or limited source-separation capability. Owing to their relatively modest angular resolution and wide fields of view, HAWC/LHAASO have lower point-source separation capability than H.E.S.S., such that multiple nearby emitters can overlap within a single observed region and contaminate the measured fluxes. However, because both HAWC and LHAASO are primarily sensitive to energies above $\sim 10$ TeV, source contamination is expected to be less severe in practice, with the contamination level estimated to be on the order of 10–20\% around 100 TeV \citep{Cao2019(LHAASO)}.

By contrast, H.E.S.S. reaches its highest sensitivity near $\sim 1$ TeV, an energy range populated by a wide variety of Galactic $\gamma$-ray emitters, including PWNe, SNRs, $\gamma$-ray binaries, and diffuse Galactic-plane emission. Flux contamination and source confusion are therefore expected to have a much stronger impact on flux measurements and source classification in H.E.S.S. observations. Indeed, some TeV sources detected by H.E.S.S. have been reported to suffer contamination levels exceeding 40\% \citep[][]{HESScollaboration2018}. Although such contamination does not necessarily produce false detections, it can hinder the identification of individual PWNe, particularly in crowded regions where multiple $\gamma$-ray sources overlap both spatially and spectrally. In such cases, emission from neighbouring sources may either mask the PWN component or make the observed morphology appear more extended and ambiguous, potentially leading to an underestimation of the true PWN population. By comparison, the higher-energy focus of HAWC/LHAASO likely mitigates this effect. Nevertheless, the combined influence of these observational limitations may contribute to the apparent differences in the inferred beaming fractions among the three observatories.

\subsection{MC population synthesis of PWN-associated pulsars}

\begin{figure*}
    \centering
    \includegraphics[width=0.95\textwidth]{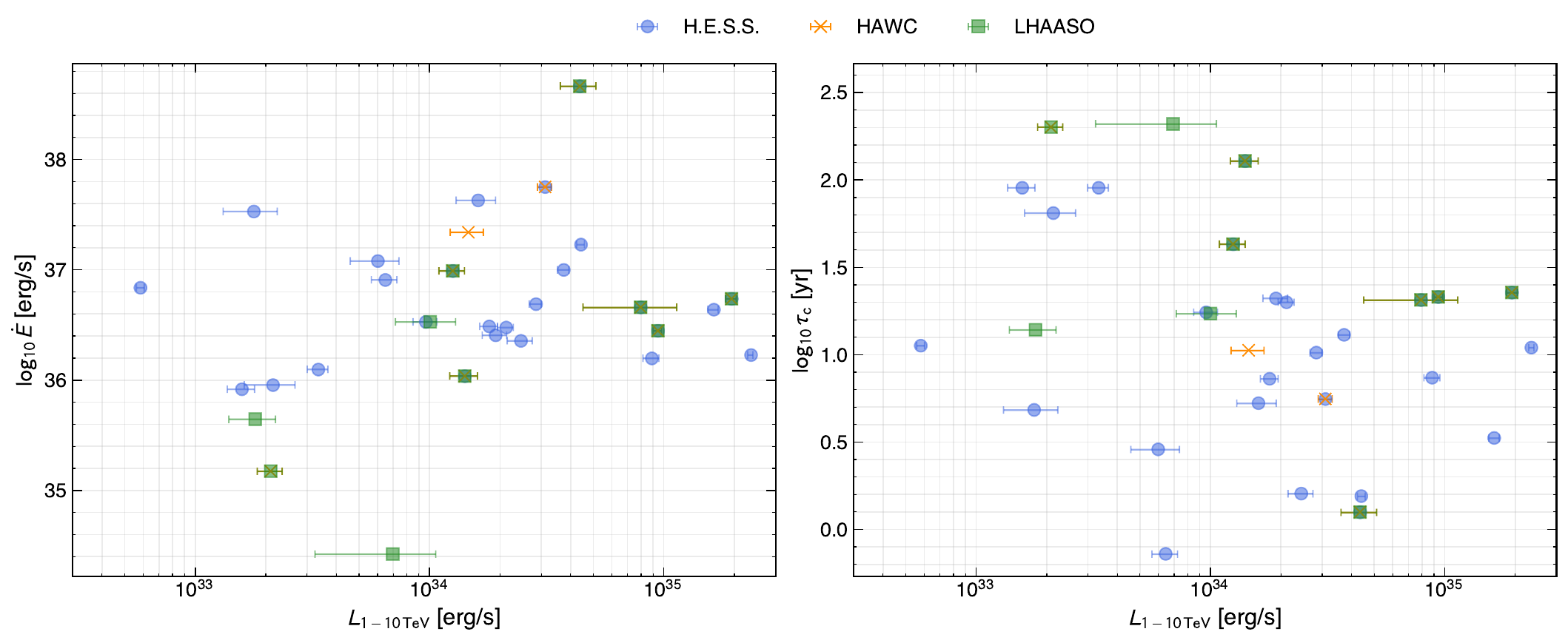}
    \caption{Comparison of the TeV luminosity of PWNe with pulsar spin-down luminosity (left) and spin-down age (right). Blue circles, orange crosses, and green squares represent PWNe detected by
H.E.S.S., HAWC, and LHAASO, respectively.
Horizontal error bars indicate the systematic uncertainties in the TeV luminosity
arising from the uncertainties in the spectral index and flux.}
    \label{fig:PWN_TeV}
\end{figure*}

\begin{table*}
\centering
\caption{Models and parameter distributions adopted as inputs for the MC population synthesis of PWN-associated pulsars.}
\label{tab:mc_inputs}
\resizebox{\textwidth}{!}{%
\begin{tabular}{l l p{0.52\textwidth} l}
\hline\hline
Quantity & Model/Distribution & Description & Reference \\
\hline
\multicolumn{4}{c}{\itshape Birth / initial properties} \\
\hline
Age (constant birth rate) & Uniform &
$t \sim \mathcal{U}(0, t_{\max})$ &
This work \\
Initial spin period, $P_{0}$ & Log-normal &
$\log_{10}(P_{0}/{\rm s}) \sim \mathcal{N}(-1.40,\,0.35^{2})$ &
\citet{Shi2024}\\
Initial dipole field, $B_{0}$ & Log-normal &
$\log_{10}(B_{0}/{\rm G}) \sim \mathcal{N}(12.４,\,0.35^{2})$ &
\citet{Shi2024}\\
Initial inclination, $\alpha_{0}$ & Isotropic &
$\cos\alpha_{0} \sim \mathcal{U}(0,1)$ &
\citet{Shi2024} \\
Birth position & Spiral arms &
Birth sites trace Galactic spiral arms &
\citet{Faucher2006,Cieslar2018} \\
Natal kick velocity & Two-
component isotropic Maxwellian distribution &
$wf_{\max}(v|\sigma_1) + (1-w)_{\max}(v|\sigma_2) $ with $w = 0.2,\, \sigma_1 = 56\,\mathrm{km/s}\, \sigma_2 = 336\,\mathrm{km/s}$ &
\citet{Igoshev2020,Shi2024} \\
Electron density (for DM) & Galactic $n_{e}$ model &
Dispersion measure computed with the electron-density model &
\citet{Yao2017} \\
\hline
\multicolumn{4}{c}{\itshape Spin / inclination evolution and Galactic motion} \\
\hline
Spin-down and alignment torque & Force-free/MHD-motivated &
$(k_{0},k_{1},k_{2})=(1,1,1)$ & \citet{Philippov2014}
\\
Galactic motion & Orbit integration &
Present-day positions from numerical integration in a Galactic potential &
\citet{Faucher2006,Cieslar2018} \\
\hline
\multicolumn{4}{c}{\itshape PWN model and TeV selection} \\
\hline

PWN luminosity (H.E.S.S.) & Log-normal &
$\log_{10}(L_{1-10}/{\rm erg\,s^{-1}})\sim \mathcal{N}(34,\,0.5^{2})$ &
This work \\
PWN luminosity (HAWC/LHAASO) & Log-normal &
$\log_{10}(L_{10-100}/{\rm erg\,s^{-1}})\sim \mathcal{N}(33.5,\,0.5^{2})$ &
This work \\
TeV selection & Integrated sensitivity &
1--10~TeV for H.E.S.S.; 10--100~TeV for HAWC/LHAASO &
This work \\
\hline
\multicolumn{4}{c}{\itshape Radio / $\gamma$-ray emission and detectability} \\
\hline
Radio luminosity & Power-law in $P,\dot{P}$ &
$(\alpha_{\rm radio},\beta_{\rm radio})=(-1,0.5)$ &
\citet{Ridley2010} \\
Radio spectrum & Power-law &
Spectral index $\alpha=-1.6$ & \citet{Jankowski2018}
 \\
Intrinsic duty cycle & Fixed &
$\delta_{\rm int}=0.05$ &
This work \\
Radio detectability & Radiometer equation &
Survey setups from \citet{Shi2024}; detected if detectable in at least one of 14 surveys &
\citet{Dewey1985,Shi2024} \\
$\gamma$-ray luminosity & Power-law in $P,\dot{P}$ &
$(\alpha_{\gamma},\beta_{\gamma})=(-1.93,0.75)$ &
\citet{Gonthier2018} \\
$\gamma$-ray detectability & Sensitivity map &
Detection threshold from the 3rd-\textit{Fermi}/LAT sensitivity map &
\citet{Smith2023(3rdFermi)} \\
\hline
\end{tabular}}
\end{table*}

\begin{table}
\centering
\caption{Adopted parameters and resulting beaming fractions
from the population synthesis for each TeV survey.}
\label{tab:mc_beaming_params}
\begin{tabular}{lcccccc}
\hline
Survey &
$t_{\max}$ [yr] &
$\rho_{\rm radio}$ [$^\circ$] &
$\rho_{\gamma}$ [$^\circ$] &
$f_{b,\rm radio}$ &
$f_{b,\gamma}$ \\
\hline
H.E.S.S.  & $1\times10^{5}$ & 30 & 30 & 0.33 & 0.24 \\
HAWC     & $3\times10^{5}$ & 10 & 5 & 0.09 & 0.06 \\
LHAASO   & $3\times10^{5}$ & 10 & 5 & 0.08 & 0.08 \\
\hline
\end{tabular}
\end{table}

To examine the factor-of-two discrepancy in the beaming fractions inferred from H.E.S.S. and HAWC/LHAASO, it would be valuable to investigate a population synthesis of pulsars associated with PWNe. Such simulations would allow us to determine which parameter choices can naturally produce a discrepancy of this magnitude. 

Here, pulsars in X-ray are not considered. 
X-ray emission generally consists of thermal emission from surface cooling and/or heated polar caps, and non-thermal emission from magnetospheric particles. Unlike in the radio and gamma-ray bands, multiple pulse components in X-rays complicate the detection of pulse profiles. Consequently, consistent detection criteria for MC simulations have not yet been established.

Our objective is to identify the geometrical conditions that reproduce the observed beaming fractions. In particular, we treat the opening angle $\rho$ and the corresponding maximum neutron star age contributing to the PWN-selected sample as free parameters. For simplicity and clarity of interpretation, we assume that the opening angle $\rho$ is the same for all neutron stars within each observational band, while allowing it to differ between bands. We further adopt a simple
top-hat beam model, in which emission is assumed to be uniform within the
opening angle and zero outside.

Our modeling framework follows \citet{Shi2024} (see also Shimasue \& Hotokezaka, in prep., for details). Below we summarise the essential outline of our approach. We first generate a neutron star population with an initial spin-period distribution($P_0$), an initial magnetic-field distribution($B_0$), and an initial magnetic-inclination distribution($\alpha_0$). NSs are then evolved with the coupled evolution equations for the spin period and the magnetic inclination angle. In parallel, we solve the NS equations of motion in the Galactic gravitational potential to obtain the present-day spatial distribution of NSs on the Galactic plane prior to applying observational selection effects. We assume that the PWN birth rate is identical to the pulsar birth rate and that every pulsar is born with an PWN. Adopting a PWN luminosity function, we then restrict the synthetic population to pulsars that would be detectable through their PWNe. For radio pulsars, we follow \citet{Shi2024} and apply the selection functions of 14 radio surveys to construct simulation samples for each survey. For $\gamma$-ray pulsars, we build mock samples in an analogous manner using the 3-rd \textit{Fermi}-LAT sensitivity map \citep{Smith2023(3rdFermi)}. The list of parameters used in our analysis is summarized in Table \ref{tab:mc_inputs} .

We describe the assumptions of the population model in more detail. We create 100000 pulsars in time until $t_{\max}$. 
The initial spin period and the initial dipole magnetic field are drawn from log-normal distributions:$\log_{10}(P_0)\sim \mathcal{N}(-1.40,0.35^2)$ and $\log_{10}(B_0)\sim \mathcal{N}(12.4,0.35^2)$ derived from \citet{Shi2024}.
Here we do not attempt to estimate the initial-condition parameters of the NS population. Our primary objective is to estimate the opening-angle and the age distribution that reproduce the observed beaming fractions. Therefore, we do not vary the initial conditions; it is sufficient to adopt a set of population parameters that reasonably reproduces the overall pulsar population.

We assume an isotropic distribution for the initial magnetic inclination angle, such that $\cos\alpha$ follows a uniform distribution. The subsequent evolution of the spin period $P$ and the magnetic inclination angle $\alpha$ between the magnetic axis and the rotation axis is modelled using spin-down prescriptions motivated by force-free or MHD simulations \citep{Philippov2014}:
\begin{align}
    \label{Pdot}\dot{P} &= \frac{4\pi^2 R_{\mathrm{NS}}^6B^2}{Ic^3P}(k_0 + k_1\sin^2\alpha)\\
    \label{alphadot}\dot{\alpha} &= -\frac{4\pi^2 R_{\mathrm{NS}}^6B^2}{Ic^3 P^2}(k_2\sin\alpha\cos\alpha),
\end{align}
where $I$ is the NS moment of inertia $I \approx 10^{45}\ {\rm g\,cm^2}$, $R_{\rm NS} = 10^6\,\mathrm{cm}$ is the NS radius, $B$ is the magnetic field strength, and $c$ is the speed of light. The parameters $k_0$, $k_1$, and $k_2$ characterise for a plasma-filled magnetosphere we set $k_0=k_1=k_2=1$. Here we consider PWN-selected pulsars which are young and we do not include the magnetic decay in solving these equations. Given the age of each simulated pulsar, we numerically solve the equations of motion in the Galactic potential to determine its present-day position. We assume that NS birth positions trace the Galactic spiral arms \citep{Faucher2006, Cieslar2018}. The natal kick velocities are drawn from a bimodal distribution \citep{Igoshev2020, Shi2024}. The electron density model in the galactic plane to calculate the dispersion measure is based on \citet{Yao2017}.

Here we adopt the TeV luminosity function of PWNe. Since the 1--10~TeV luminosity shows no clear dependence on present pulsar parameters such as $\dot{E}$ and $\tau_{\rm c}$ \citep[Figure~\ref{fig:PWN_TeV}, also see][]{Mattana2009}
, we do not parameterize the TeV luminosity directly by these quantities. Instead, we adopt a phenomenological log-normal luminosity function. This choice is also motivated by spectral-evolution studies suggesting that the TeV properties of PWNe depend primarily on the total injected rotational energy and the age of the system; because the injected rotational energy is set mainly by the birth spin period $P_0$ \citep[][]{Tanaka2013}, which is broadly utilized as log-normal distribution \citep[e.g.][]{Shi2024}. 
Accordingly, we assume
$\log_{10}(L_{1-10}/{\rm erg\,s^{-1}})\sim \mathcal{N}(34,0.5^2)$
for H.E.S.S. and
$\log_{10}(L_{10-100}/{\rm erg\,s^{-1}})\sim \mathcal{N}(33.5,0.5^2)$
for HAWC/LHAASO, the latter being motivated by the observed luminosity ratios between the H.E.S.S. and HAWC/LHAASO bands.

We adopt the integrated sensitivity in the 1--10~TeV band for H.E.S.S. and in the 10--100~TeV band for HAWC/LHAASO (see Fig.~\ref{fig:sensitivity}). Using these sensitivities, we construct volume-limited PWN samples and apply the radio and $\gamma$-ray detection thresholds.

We introduce the intrinsic radio luminosity function for radio and $\gamma$-ray as a function of $P$ and $\dot{P}$ :
\begin{align}
    L_{\nu} = f_\nu P^{\alpha_\nu}\dot{P}^{\beta_\nu}.
\end{align}
Observationally the luminosity function does not depend on the observable parameters and here we adopt the lowest power-law index to $P$ and $\dot{P}$ taken from previous study, $\alpha_\mathrm{radio} = -1$ and $\beta_\mathrm{radio}
= 0.5$ derived from \citet{Ridley2010}. For $\gamma$-ray, we applied $\alpha_{\gamma} = -1.93$ and $\beta_{\gamma} = 0.75$ derived from \citet{Gonthier2018}.  Radio flux density is reconstructed with assuming power-law spectrum with photon index $\Gamma = -1.6$ and we obtain : 
\begin{align}
    S_{\mathrm{mean}}(\nu_0) = \frac{\delta_{\mathrm{int}} L_\mathrm{radio}{}}{\Omega d^2} \frac{\Gamma + 1}{\nu_0}\left[\left(\frac{\nu_2}{\nu_0}\right)^{\Gamma+1} - \left(\frac{\nu_1}{\nu_0}\right)^{\Gamma+1}\right]^{-1},
\end{align}
where $\nu_1 , \nu_2$ is the frequency range of radio survey, $\nu_0$ is the central frequency, $\delta_{\mathrm{int}}$ is the intrinsic duty cycle, $\Omega$ is the solid angle. We fix the intrinsic duty cycle for all pulsars to the peak value, $\delta_{\mathrm{int}}=0.05$ since the observed duty-cycle distribution shows little dependence on $P$ and $\dot{P}$. 
We then determine whether a pulsar is detectable by comparing the radio flux density limit computed from the radiometer equation \citep{Dewey1985}. The survey configurations adopted in the radiometer equation are taken from \citet{Shi2024}. 
We consider 14 radio surveys, and a pulsar is counted as detected if it is detectable in at least one of them. For $\gamma$-ray, the detection limit is derived from the 3-rd \textit{Fermi}-LAT sensitivity map \citep{Smith2023(3rdFermi)}.

Using the procedure described above, we construct a simulated sample by applying the detection limits of radio or $\gamma$-ray surveys to the PWN-selected pulsars in the simulation. We denote the resulting number of detected objects as $N_{\nu,\mathrm{detect}}$. The simulated beaming fraction is then evaluated as
\begin{align}
f_{b,\nu,\mathrm{sim}} = \frac{N_{\nu,\mathrm{detect}}}{N_{\mathrm{PWN,\mathrm{detect}}}} , 
\end{align}
where  is the number of sources detectable in the radio (or $\gamma$-ray) band under the adopted survey sensitivities. We estimate the opening angle $\rho$ that reproduces the beaming fraction inferred from the TeV-selected sample in this study.

\begin{figure*}
    \centering
    \includegraphics[width=0.95\textwidth]{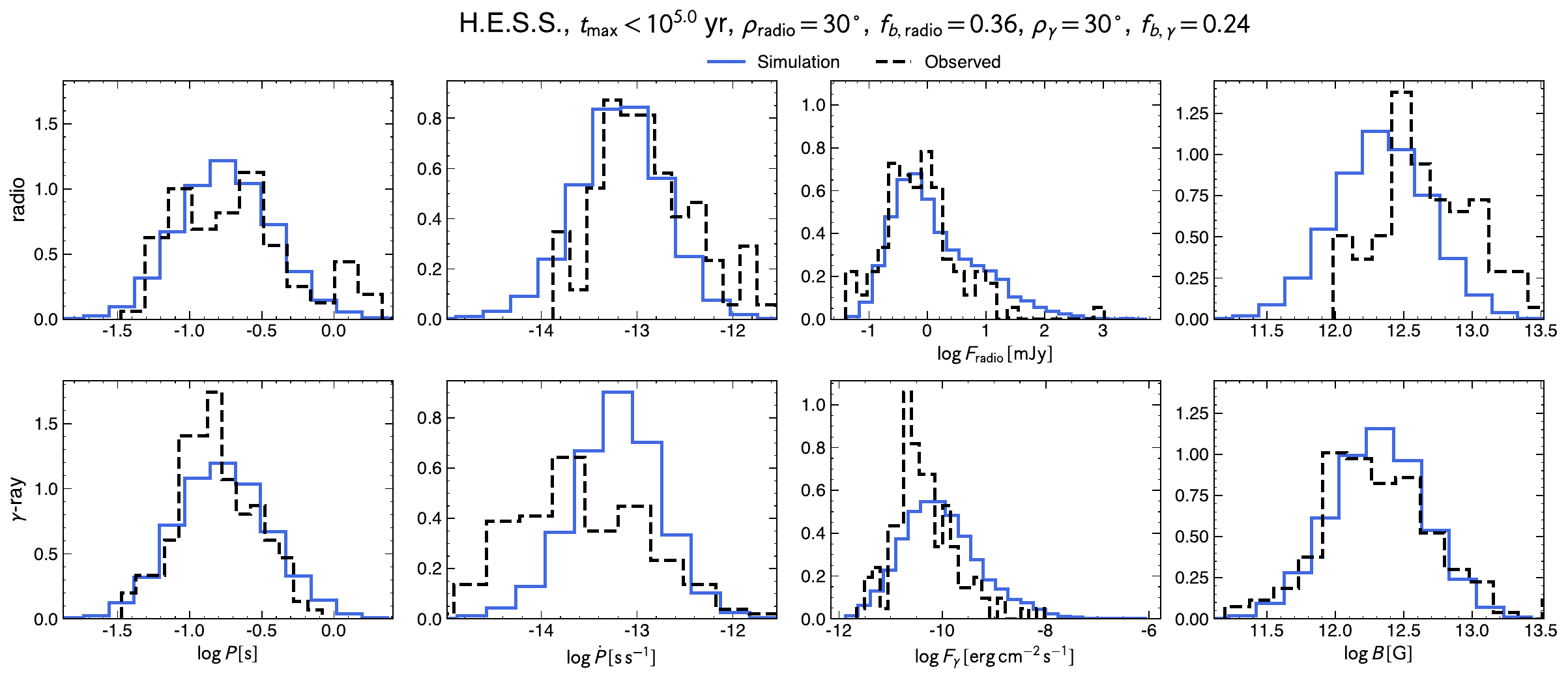}
    \caption{Comparison of the simulated (blue solid lines) and observed (black dashed lines) distributions
of the spin period $P$ (left), period derivative $\dot{P}$ (centre left),
flux (centre right), and dipole magnetic field strength $B$ (right)
for pulsars associated with PWNe detected under the H.E.S.S. observational setup,
based on the initial parameters listed in Table.
The top (bottom) panels correspond to radio ($\gamma$-ray) pulsars based on ATNF catalog \citep{Manchester2005} and the 3-rd \textit{Fermi}-LAT $\gamma$-ray pulsar catalog \citep{Smith2023(3rdFermi)}.}
    \label{fig:HESS}
\end{figure*}

\begin{figure*}
    \centering
    \includegraphics[width=0.95\textwidth]{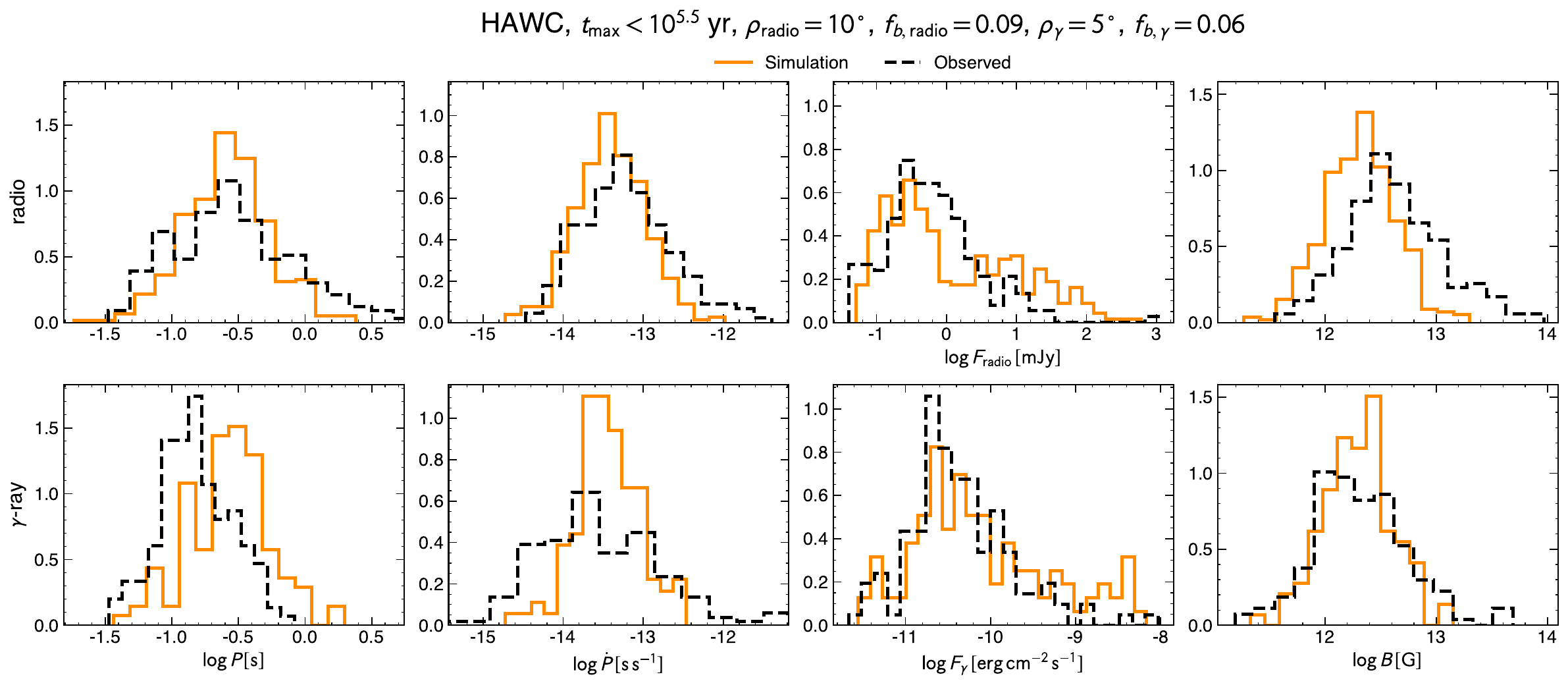}
    \caption{Similar to Fig. \ref{fig:HESS}, but HAWC set-up.}
    \label{fig:HAWC}
\end{figure*}

\begin{figure*}
    \centering
    \includegraphics[width=0.95\textwidth]{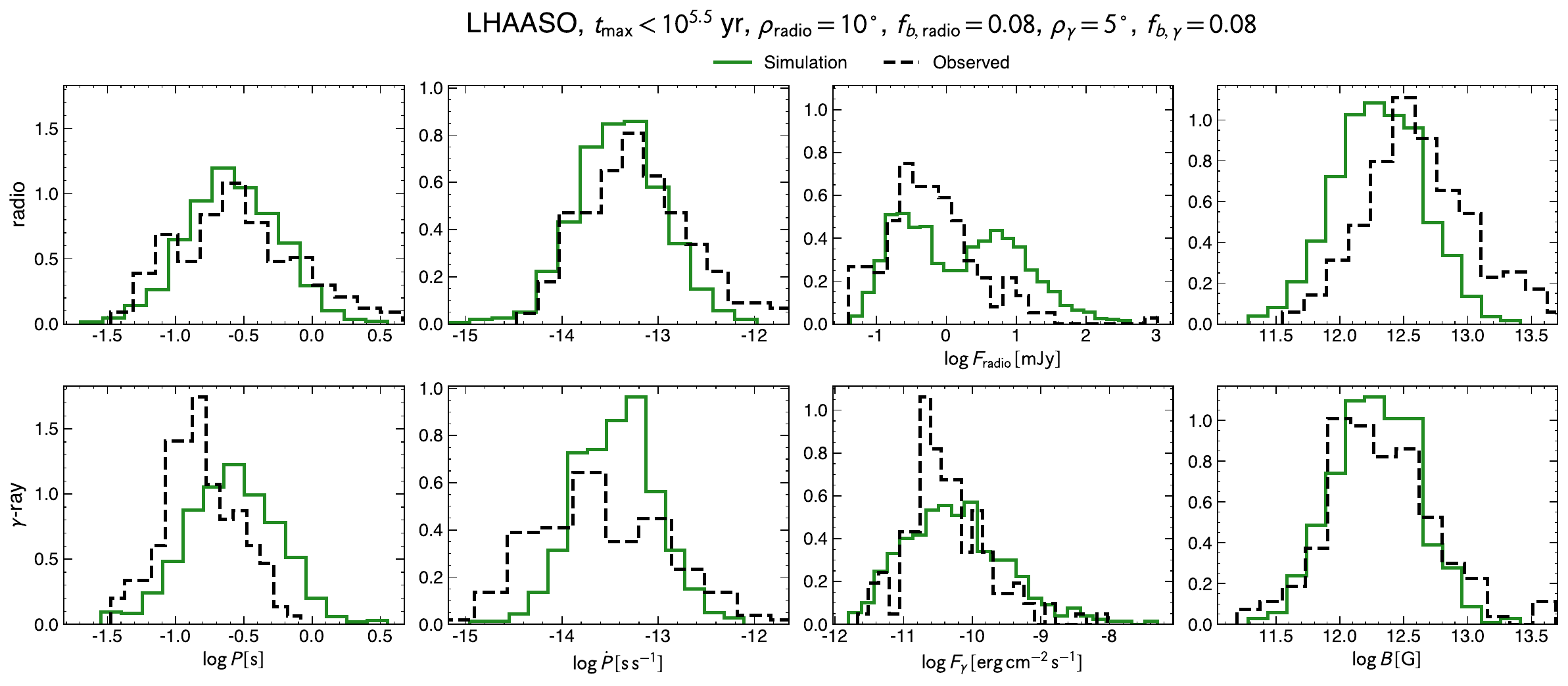}
    \caption{Similar to Fig. \ref{fig:HESS}, but LHAASO set-up.}
    \label{fig:LHAASO}
\end{figure*}

Figures~\ref{fig:HESS}, \ref{fig:HAWC} and \ref{fig:LHAASO} presents a comparison between the simulated and observed
distributions of pulsars associated with PWNe detected by H.E.S.S., HAWC/LHAASO, respectively. We find a possible parameter set of the opening angle $\rho$ and the maximum pulsar age listed in Table~\ref{tab:mc_beaming_params}. 
Specifically, the H.E.S.S. sample is reproduced by adopting relatively
large opening angles and a younger pulsar population, whereas the
beaming fractions inferred for HAWC/LHAASO require smaller opening
angles together with a larger value of $t_{\max}$. 

Considering the difference in effective volume inferred from our MC results, the beaming fraction for H.E.S.S. is 2-4 times larger than that for HAWC/LHAASO, even for the same beam opening angle. In addition,  fraction of faint neutron stars remains undetected despite their beams intersecting our line of sight, amounting to $\sim 0.4$ for H.E.S.S. and $\sim 0.6$ for HAWC/LHAASO. To explain an approximately factor-of-two to four difference in the beaming fraction, a difference in the opening angle by a factor of about 3–6 is required, implying that the opening angle for HAWC/LHAASO is systematically smaller than that of H.E.S.S..

If the opening angle decreases with pulsar age in an approximate sense, this implies that pulsars associated with PWNe detected by HAWC/LHAASO are systematically older than those associated with H.E.S.S. PWNe. 
This interpretation is also broadly consistent with the picture discussed in Sec \ref{subsec:Instrument} in terms of angular resolution and energy band. 

\begin{figure}
    \centering
    \includegraphics[width=0.95\linewidth]{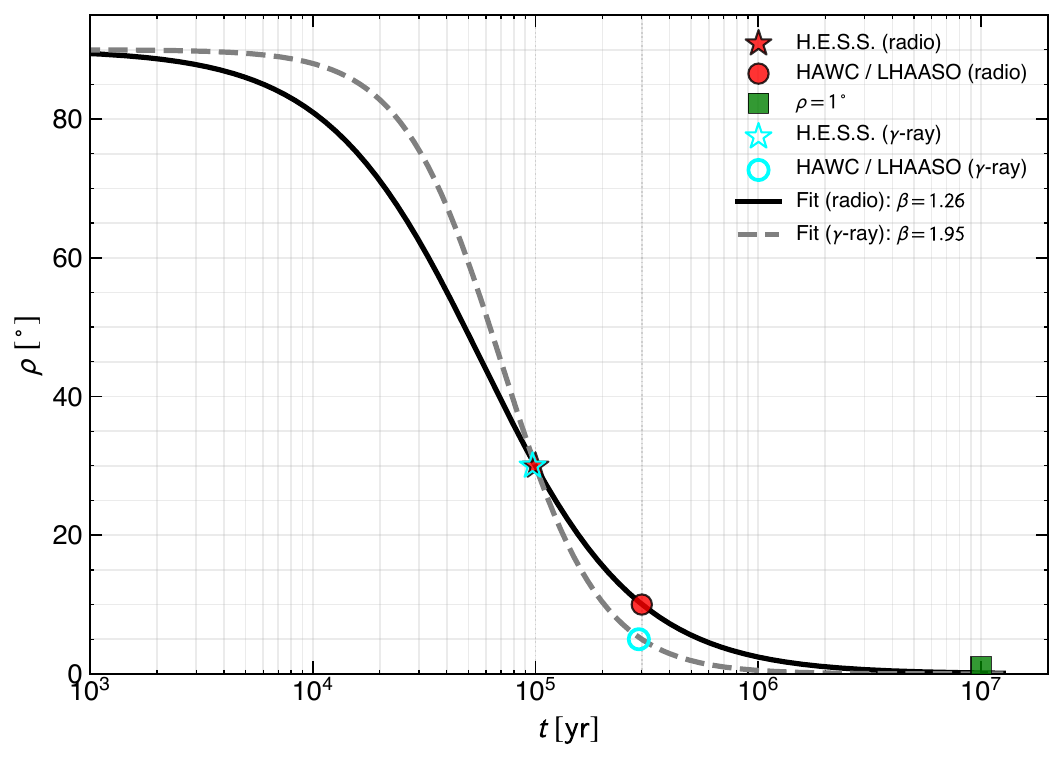}
    \caption{Time evolution of the effective radio beam opening angle $\rho$ inferred from this study.
The star and circle indicate the effective opening angles associated with the H.E.S.S. and HAWC/LHAASO samples, respectively.
The line shows a power-law fit of the form $\rho(t)\propto t^{-\beta}$, with $\beta\simeq 1.3$ (radio) and $2$ ($\gamma$-ray).} 
    \label{fig:fit}
\end{figure}

\begin{figure*}
    \centering
    \includegraphics[width=0.95\textwidth]{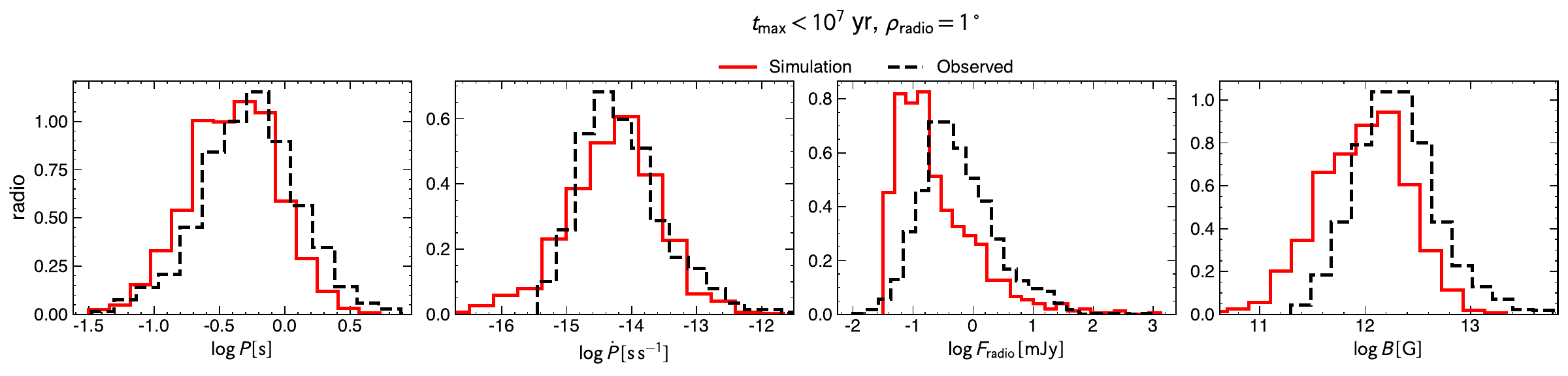}
    \caption{Similar to Fig.~\ref{fig:HESS}, but showing the radio pulsar population younger than $10^7\,\mathrm{yr}$, without imposing any TeV-survey selection. }
    \label{fig:radio}
\end{figure*}

To this end, we adopt a single pulsar population with a time-dependent opening angle $\rho(t)$ and demonstrate that the beaming fractions inferred from TeV-selected samples and the statistical properties of the radio pulsar population reported in earlier studies can be simultaneously reproduced within a unified framework.
We explicitly show that at least one representative set of parameters provides a self-consistent explanation of both without introducing any mutual inconsistency.
As a concrete example, we present a representative parameter set that reproduces the effective radio opening angles inferred for all three TeV surveys.
Specifically, we adopt an opening-angle evolution of the form $
\rho(t)=\frac{90^\circ}{1+(t/t_{\rm c})^\beta}$,
which approaches $90^\circ$ at $t=0$.
The best-fit parameters are $(t_{\rm c},\beta)=(5.75\times10^4\,\mathrm{yr},\,1.26)$ for radio and $(7.01\times10^4\,\mathrm{yr},\,1.95)$ for $\gamma$-ray (Figure~\ref{fig:fit}).
Figure~\ref{fig:radio} shows that, even when the entire radio pulsar population younger than $10^7\,\mathrm{yr}$ is taken into account, the simulated distributions remain consistent with the observed distribution.
This suggests that the results of this study are consistent with the statistical properties of the observed radio pulsar population.

\subsection{Emission geometry of pulsars for each band}


The cumulative beaming fractions in the radio, $\gamma$-ray bands are broadly consistent with previous studies \citep{Tauris1998,Johnston2020,Turner2025}. 
The results of our MC simulations suggest that the opening angle of radio and $\gamma$-ray may decrease on a timescale of $\sim 10^5$ yr. The rotational phases of the intensity peaks in $\gamma$-ray and radio emission are generally different; these emissions are thought to originate from distinct regions \citep[][]{Smith2023(3rdFermi)}. However, the similarity implies that the radio, $\gamma$-ray and X-ray emission regions in pulsars are related, showing a similar opening angle evolution.

For $\gamma$-ray and X-ray, we first consider a scenario in which the emission arises from the current sheet region outside the light cylinder. In this picture, the opening angle strongly depends on the magnetic inclination angle between the magnetic axis and the rotation axis. 
Here we have already adopted the magnetic inclination angle alignment based on Eqs.~(\ref{Pdot}) and (\ref{alphadot}). Therefore, the decrease in the opening angle may be attributed to the additional alignment mechanisms \citep[for a review, see][]{Li2023}.  

In the outer gap model, the spatial scale of the gap is primarily determined by the mean free path of $\gamma$-rays \citep{Cheng1986}, which increases with age. The region where particles are predominantly supplied would correspond to the upper boundary of the gap. As this boundary moves farther from the last open field lines, the opening angle becomes smaller \citep{Watters2009, Kisaka2011}. Our results may therefore reflect the evolution of the $\gamma$-ray mean free path, although, as in the current sheet scenario, an additional magnetic axis alignment may also play a role.

For radio emission, the evolution of the opening angle on a timescale of $\sim 10^5$ yr appears to be faster than that expected from the evolution of the polar cap size $ \rho \propto P^{-1/2} \propto t^{-1/4}$ under the assumption of constant magnetic field strength and magnetic inclination angle. This suggests such as magnetic inclination angle alignment and changes in the current structure within the polar cap.

\section{CONCLUSIONS}\label{sec:conclusion}

We have presented a largely geometry-independent estimate of the pulsar beaming fraction using TeV-selected Galactic PWNe and Unid TeV sources based on TeVCAT ver2.2 \citep{Tevcat2008}.
Assuming that TeV emission from PWNe is approximately isotropic and that Unid TeV sources in the Galactic plane are dominated by PWNe whose pulsar beams do not intersect our line of sight, we quantified the beaming fraction as the number ratio, and evaluated it separately for three TeV surveys (H.E.S.S., HAWC, and LHAASO) and for pulsars detected in radio, X-rays, and $\gamma$-rays based on ATNF catalog \citep{Manchester2005} and 3-rd \textit{Fermi}-LAT  catalog \citep{Smith2023(3rdFermi)}.

Our main findings are as follows.
\begin{enumerate}
    \item Within each survey, the inferred cumulative beaming fractions show no strong dependence on the pulsar observing band.
    For H.E.S.S., we obtain cumulative beaming fraction, $f_b \simeq 0.23$--$0.36$ with $\dot{E}_{\mathrm{th}} = 10^{36}\,\mathrm{erg/s}$ or $\tau_{c\mathrm{,\,th}} = 10^5\,\mathrm{yr}$, while for HAWC/LHAASO we obtain systematically smaller values, $f_b \simeq 0.04$--$0.13$ with $\dot{E}_{\mathrm{th}} = 10^{34.5}\,\mathrm{erg/s}$ or $\tau_{c\mathrm{,\,th}} = 10^{5.5}\,\mathrm{yr}$ (Table~\ref{tab:beam}).
    \item The similarity of $f_b$ among radio, $\gamma$-ray, and X-ray pulsars within the same survey suggests that, for young pulsars associated with TeV PWNe, the emission geometries across bands may be more related each other.
    This motivates a re-assessment of models in which radio beams are systematically much narrower than $\gamma$-ray beams.
    \item The factor of 2--4 discrepancy between H.E.S.S. and HAWC/LHAASO likely reflects survey-dependent selection effects.
    In particular, wide-field of view and broad point spread function are more sensitive to older and more extended PWNe \citep{HESS2018PWN, Alfaro2024(HAWC), Cao2019(LHAASO), Cao2024}, and their PWN classifications rely more strongly on previous identifications.
    These differences can naturally bias the PWN/Unid number ratio and hence the inferred $f_b$.
\end{enumerate}

The forthcoming CTAO will significantly advance this situation \citep{CTAO2019, CTAO2023}.
With its order-of-magnitude improvement in flux sensitivity compared to existing instruments such as H.E.S.S., HAWC, and LHAASO, together with its superior angular and energy resolution, CTAO will be capable of resolving complex source morphologies and distinguishing between overlapping emitters even in crowded regions of the Galactic plane\footnote{\href{https://www.ctao.org/for-scientists/performance/}{https://www.ctao.org/for-scientists/performance/}}.
At 1 TeV, CTAO will achieve an angular resolution of approximately $0.06^{\circ}$, the finest among any existing or planned ground-based $\gamma$-ray observatories, allowing for precise source localization and robust discrimination from nearby emitters.
Moreover, the contamination level due to adjacent sources is expected to be as low as $\sim$10 \% at 1 TeV, while the residual cosmic-ray background will be suppressed to below 1 \% of the total event rate.
These capabilities will dramatically reduce the impact of source confusion and enable the detection of a much larger population of faint or previously unresolved PWNe.
Future CTAO observations will therefore play a crucial role in establishing a unified picture of pulsar emission geometry and refining our understanding of the relationship between PWNe and their associated pulsars.

\section*{DATA AVAILABILITY}

The data that support the findings of this study are
available within the article. The code and csv files are available at \href{https://github.com/Takumi-Shima-sue/beaming_fraction.git}{https://github.com/Takumi-Shima-sue/beaming-fraction.git}

\section*{Acknowledgements}

We thank Kenta Hotokezaka for the helpful conversation. We also appreciate Hidetoshi Kubo for providing the sensitivity curve of CTAO. This work was supported by JSPS KAKENHI Grant Numbers JP22K03681, JP26K07066 (SK, SS), and JP23K22538 (SK).

\bibliographystyle{mnras}
\bibliography{main}





\bsp	
\label{lastpage}
\end{document}